\documentclass[journal,letterpaper]{IEEEtran}
\IEEEoverridecommandlockouts

\usepackage[linesnumbered,ruled,vlined]{algorithm2e}

\SetCommentSty{mycommfont}
\usepackage{amsmath}
 \usepackage{tcolorbox}
 \usepackage{balance}
\usepackage{xcolor}
\usepackage{multirow}
\usepackage{enumitem}
\usepackage{color, colortbl}
\usepackage{amssymb}
\usepackage{subcaption}

\usepackage{soul,comment}
\specialcomment{review}{\color{black}}{\color{black}}
\specialcomment{review2}{\color{black}}{\color{black}}
\usepackage{hyperref}
\hypersetup{breaklinks=true}
\usepackage{multicol}
\usepackage{amsmath}
\usepackage{amsfonts}
\usepackage{tikz}
\usetikzlibrary{shapes}
\usepackage{adjustbox}
\usepackage[T1]{fontenc}
\usetikzlibrary{decorations.pathmorphing}
\definecolor{eclipseBlue}{RGB}{42,0.0,255}
\definecolor{eclipseGreen}{RGB}{63,127,95}
\definecolor{eclipsePurple}{RGB}{127,0,85}
\definecolor{Gray}{gray}{0.9}
\newcommand{\coldscr}{\cellcolor{Gray}}
\usepackage{color}
\usepackage{rotating}
\colorlet{BLUE}{blue}
\usepackage{colortbl}
\definecolor{LightCyan}{RGB}{255, 255, 100}
\definecolor{LightCyan1}{RGB}{162, 255, 100}

\newcommand\systemname{\textsc{TEL-C}\space}
\newcommand\systemnameDC{\textsc{TEL-D}\space}
\newcommand\systemnameDCTEL{\textsc{TEL-D}}
\newcommand\systemnameTEL{\textsc{TEL-C}}
\usepackage{listings}

\lstdefinelanguage{p4}
{
	morekeywords={
		action, apply, bit, bool, const, control, default, else, enum, error, 
		extern, exit, false, header, if, in, inout, int, match_kind, package, 
		parser, out, return, select, state, struct, switch, table, transition, 
		true, tuple, typedef, varbit, verify, void,
	},
	sensitive=true, % keywords are case-sensitive
	morecomment=[l]{//}, % l is for line comment
	morecomment=[s]{/*}{*/}, % s is for start and end delimiter
	morestring=[b]" % defines that strings are enclosed in double quotes
}
\usepackage{mathtools}
\DeclarePairedDelimiter\ceil{\lceil}{\rceil}

\tikzstyle{mybox} = [draw=black, fill=white,  thin, rectangle, rounded corners, 
inner sep=10pt, inner ysep=10pt]
\tikzstyle{fancytitle} =[text=white, font=\bfseries]

\usepackage{booktabs} % For formal tables

\begin{document}
%\bstctlcite{IEEEexample:BSTcontrol}

\title{ TEL: Low-Latency Failover Traffic Engineering in Data Plane}

\author{
 \IEEEauthorblockN{
 Habib Mostafaei\IEEEauthorrefmark{1},
 Mohammad Shojafar\IEEEauthorrefmark{2},	
 Mauro Conti\IEEEauthorrefmark{3}\\
 }
  \IEEEauthorblockA{\IEEEauthorrefmark{1}
  Technische Universit\"at Berlin,
 }
 \IEEEauthorblockA{\IEEEauthorrefmark{2}
  University of Surrey,
 }
 \IEEEauthorblockA{\IEEEauthorrefmark{3}
  University of Padua\\
 }
 habib@inet.tu-berlin.de,  m.shojafar@surrey.ac.uk, conti@math.unipd.it
% 	\thanks{
% 	}
}

\maketitle

\begin{abstract}
Modern network applications demand low-latency traffic engineering in the 
presence of network failure, while preserving the quality of service 
constraints like delay and capacity. Fast Re-Route (FRR) mechanisms are widely 
used for traffic re-routing purposes in failure scenarios. Control plane FRR 
typically computes the backup forwarding rules to detour the traffic in the 
data plane when the failure occurs. This mechanism could be computed in the 
data plane with the emergence of programmable data planes.

In this paper, we propose a system (called \texttt{TEL}) that contains 
\textit{two} FRR mechanisms, namely, \systemname and \systemnameDCTEL. The 
first one computes backup forwarding rules in the control plane, satisfying 
max-min fair allocation. The second mechanism provides FRR in the data plane. 
Both algorithms require minimal memory on programmable data planes and 
well-suited with modern line rate match-action forwarding architectures (e.g., 
PISA). We implement both mechanisms on P4 programmable software switches (e.g., 
BMv2 and Tofino) and measure their performance on various topologies. The 
obtained results from a datacenter topology show that our FRR mechanism can 
improve the flow completion time up to 4.6x–7.3x (i.e., small flows) and 
3.1x–12x (i.e., large flows) compared to recirculation-based mechanisms, such 
as F10, respectively.
\end{abstract}

\begin{IEEEkeywords}
Traffic engineering, network monitoring, programmable data plane, low-latency, 
link failure, reinforcement algorithm.
\end{IEEEkeywords}

\section{Introduction}\label{sec:1}

Recent cloud datacenters run numerous applications on their networks that are 
interconnected through several servers. The applications have low-latency 
requirements and demand fast rerouting in the case of any failure. To address 
these requirements, Fast Re-Route (FRR) mechanisms are widely used to reroute 
the traffic~\cite{PURR-conext2019}. Control plane FRR typically computes the 
backup forwarding rules to detour the traffic in the data plane when the 
failure occurs. This proactive way of maintaining the backup forwarding rules 
in the switches improves the network robustness and availability.

The control plane FRR solutions are widely adopted by the network equipment 
vendors. Such solutions allow the network administrators to implement the 
network functionalities as a black-box option~\cite{PURR-conext2019}. 
Nevertheless, these mechanisms provide less flexibility to the network 
operators for customized packet processing. Recent advances in programmable 
data planes~\cite{p4-2014} offer flexible packet header processing, which is 
useful in many use-cases, including network monitoring~\cite{pint-sigcomm20}, 
and traffic load balancing~\cite{hula-SOSR16} to state few examples. Packet 
recirculation can be used as a simple mechanism in the programmable pipelines 
to detour the traffic to the pipeline's input port. Then, the pipeline can 
select the different egress port when the failure is detected. However, this 
solution decreases the packet processing throughout and increases its 
latency~\cite{PURR-conext2019}.

\begin{review2}
	Some business networking applications have stringent Quality of Service 
	(QoS) requirements such as latency and bandwidth. Control plane FRR 
	solutions could satisfy the QoS requirements by using mechanisms such as 
	max-min fair allocation~\cite{jose2019distributed,max-min-CCR18}. In 
	contrast, data plane solutions are fast in rerouting the traffic but 
	unaware of QoS constraints dictated by the traffic policies. The reason is 
	that the selected nodes in the path may not have enough capacity to steer 
	the traffic. 
\end{review2}

\subsection{Motivations}\label{sec:1.1}
\begin{review2}
	In literature, several schemes exist to deal with link failures in 
	programmable data 
	plane~\cite{PURR-conext2019,markopoulou2008characterization,DDC-NSDI13,qu2019sqr,holterbach2019blink,qian2019flexgate,F10-NSDI13}.
	 For instance, F10~\cite{F10-NSDI13} recirculates the traffic of failed 
	ports until an alternative port is explored. However, packet recirculation 
	has a low performance~\cite{PURR-conext2019}. Besides, the authors 
	in~\cite{PURR-conext2019} create a set of new FRR primitives and implement 
	them in P4~\cite{p4-2014} to preserve high availability and low latency.  
	Blink~\cite{holterbach2019blink} is a fast data-driven remote failures 
	algorithm to deal with inter-domain failures in P4. It tracks failure 
	signals and monitors the link rate to reroute the traffic automatically. 
	\emph{FlexGate}~\cite{qian2019flexgate} proposes a rule placement algorithm 
	to mitigate the link failure on various network functions at high 
	throughput. Nevertheless, none of these approaches simultaneously consider 
	the control plane and data plane FRR mechanisms. Motivated by these 
	considerations, our intention in this paper is to propose \textit{two} FRR 
	mechanisms that implement on control- and data planes simultaneously. The 
	first one is a control plane-based mechanism that preserves QoS 
	requirements using max-min fair allocation (i.e., traffic engineering). The 
	second one provides low-latency FRR in the data plane.  Both mechanisms 
	require minimal memory on programmable data planes and well-suited with 
	modern line rate match-action forwarding architectures (e.g., PISA). 
\end{review2} 
Specifically, we respond to the following questions: \textit{i)} Is it possible 
to provide failover traffic engineering? (see Section~\ref{sec:model}) 
\textit{ii)} How can we preserve a set of QoS constraints in the steering 
traffic of different users? (see Section~\ref{sec:path-selection}) And, 
\textit{iii)} How can we solve the max-min fair allocation problem in a linear 
time? (see Section~\ref{sec:path-selection}).

\subsection{The goal of the paper and contributions}\label{sec:1.2}
% \textcolor{teal}{I agree with the reviewer said that the goal is not clear! 
%We never said that all these blah blah is to keep the FCT of flows low. We 
%should emphasis this point!}

\begin{review}
	We propose \textit{two} FRR mechanisms, i.e., one in the control plane and 
	one in the data plane, to detour the traffic in the case of failure. The 
	control plane solution, \systemnameTEL, can satisfy max-min fair allocation 
	in assigning flows to the network links. This solution is a proactive 
	solution and should be executed before network operations. To accomplish 
	this, we use the Distributed Learning Automaton (DLA) to explore the paths 
	while considering multiple QoS constraints, such as delay and capacity of 
	the link. This problem is known to be an NP-hard 
	problem~\cite{qos-routing-jsac96}, and our approach finds each candidate 
	solution in linear time. It is an iterative approach to find the best 
	optimal path among all possible paths.
	
	The second solution, \systemnameDCTEL, is similar to the state-of-the-art 
	data plane solutions. In this solution, we select a random egress port from 
	the available list of ports to reroute the traffic when a failure occurs. 
	It is well-suited in datacenter topologies.

	The proposed solutions avoid recirculations and lead to the following goals:
	\begin{itemize} [leftmargin=*]
		
		\item \textit{Low latency and high throughput:} In case of failure, the 
		traffic will rapidly reroute to the next available port without 
		performance degradation and irrespective of the number of failures. 
		
		\item \textit{Memory cost}: Our proposed FRR mechanisms occupy minimal 
		memory on the P4-enabled devices. 
		
		\item \textit{Flexibility}: The proposed FRR mechanisms could guarantee 
		a set of link failures. It provides two sets of primary and backup 
		forwarding rules.
		
	\end{itemize}
	
	Hence, we summarize our main contributions as follows:
	\begin{itemize}[leftmargin=*]
		\item We formalize the traffic engineering with QoS requirements as a 
		max-min fair allocation problem. 
		\item To solve it, we propose a control-based FRR mechanism using a 
		reinforcement learning algorithm that selects shortest paths using DLA, 
		which finds primary and backup paths for each traffic demand of each 
		network application/service. %and % for implementing a low-latency 
		%%failover traffic engineering in P4. We assume that have $k$ 
		%%applications demanding $k$ unique paths.
		\item We propose a data plane FRR mechanism that can be used as an FRR 
		primitive in programmable data planes. 
		\item We checked the feasibility of both FRR mechanisms in P4 and BMv2 
		and Tofino software switches.
		\item Finally, we evaluate our solutions on various topologies. To be 
		precise, we compare \systemname against Yen's K-shortest path algorithm 
		on the length of shortest paths, the algorithms' running time and 
		traffic load. Also, our evaluation on a datacenter topology shows that 
		\systemnameDC can improve the Flow Completion Time (FCT) up to 
		4.6x–7.3x (i.e., small flows) and 3.1x–12x (i.e., large flows) compared 
		to recirculation-based mechanisms, such as F10~\cite{F10-NSDI13}, 
		respectively. 
	\end{itemize}
	
\subsection{Roadmap}\label{sec:1.3}
	%\noindent\textbf{Organization.} 
We organize the paper below. Section~\ref{sec:model} describes the system 
	model explaining the optimization problem. In Section~\ref{sec:algo}, we 
	present our FRR mechanisms. Section~\ref{sec:proof} presents the 
	proof-of-concept in P4-enabled devices. The simulation results are 
	presented in Section~\ref{sec:results}.  We explain the practical appliance 
	of our proposed FRR mechanisms in Section~\ref{sec:future}. 
	The related work comes in Section~\ref{sec:relatedwork}.
	Finally, we conclude our work in Section~\ref{sec:conclusion}.
\end{review}

\section{System Model}\label{sec:model}

\begin{table*}[!htpb]
	\centering
	\scriptsize
	\caption{\small Main notation.}
	\label{tab:tab1}
	%\rowcolors{2}{gray!25}{white}
	\resizebox{\textwidth}{!}{
		\begin{tabular}{|p{0.37cm}|c|p{7.7cm}|p{2.3cm}|p{1.6cm}|}
			\hline
			\coldscr \textbf{Type} & \coldscr \textbf{Symbol} & \coldscr 
			\textbf{Definition} & \coldscr \textbf{Type - Unit} & \coldscr 
			\textbf{Appears in Eq.} \\
			\hline
			%\begin{tabular}{|c|p{6cm}|}
			\multirow{3}{0.7cm}{\begin{sideways}\textbf{Set}\end{sideways}}& 
			$\mathcal{S}$ & Set of P4 switches, where $\mid \mathcal{S}\mid= 
			S$&-&-\\
			&$\mathcal{E}$ & Set of edges, where $\mid \mathcal{E}\mid= 
			E$&-&-\\ 
			&$\mathcal{P}$&Set of paths&-&-\\
			&$\mathcal{F}$&Set of flows&-&-\\
			\hline \hline
			\multirow{3}{0.7cm}{\begin{sideways}\textbf{Index}\end{sideways}}& 
			$f$ & Index of flow, $f\in \mathcal{F}$&Integer - [units]&-\\
			&$n$ & Index of node/switch, $n\in \mathcal{S}$&Integer - 
			[units]&-\\ 
			&$m$& Index of node/switch, $m\in \mathcal{S} $&Integer - 
			[units]&-\\
			\hline \hline
			&$R^f$& Required bandwidth for flow $f$&Continuous - 
			[bps]&\eqref{eq1}\\
			&$\mu$& Crossing traffic ratio to total bandwidth of each 
			link&Continuous - [units]&\eqref{eq1}\\
			\multirow{5}{0.7cm}{\begin{sideways}\textbf{Input 
			Parameters}\end{sideways}}  &$B_{(n,m)}$& Matrix of link bandwidth 
			between switch $n$ and $m$&Continuous - [bps]&\eqref{eq1}\\
			&$\Phi_{(n,m)}^f$&Network resource assignment matrix between switch 
			$n$ and $m$ for flow $f$&Binary - 
			[units]&\eqref{eq1},\eqref{eq2},\eqref{eq3},\eqref{eq4}\\
			&$s^{f}$&Source switch for flow $f$&Continuous - 
			[units]&\eqref{eq2}\\
			&$d^{f}$&Destination switch for flow $f$&Continuous - 
			[units]&\eqref{eq2}\\
			&$T^p$& Maximum tolerable delay for path $p$&Continuous - 
			[ms]&\eqref{eq3}\\
			&$D_{(n,m)}$& Propagation delay between switch $n$ and 
			$m$&Continuous - [ms]&\eqref{eq3},\eqref{eq:objective}\\
			&$B^u(n,m)$& Bandwidth usage between switch $n$ and $m$&Continuous 
			- [bps]&\eqref{eq:objective}\\
			&$C^f_{(n,m)}$ & Cost of flow $f$ while crossing switch $n$ and 
			$m$&Continuous - [units]&\eqref{eq:cost}\\
			&$C_{(n,m)}$& Steering traffic cost between switch $n$ and 
			$m$&Continuous - [units]&\eqref{eq:cost}, \eqref{eq:objective}\\
			&$\alpha$,$\lambda$, $\zeta$& Coefficients&Continuous - 
			[units]&\eqref{eq:objective}\\
			\hline\hline
			\multirow{2}{0.7cm}{\begin{sideways}\textbf{ Var}\end{sideways}} & 
			$\phi_{(n,m)}$ & Network resource assignment matrix between node 
			$n$ and $m$ &Continuous - [units]&\eqref{eq:objective}\\ 
			& $p$ & A path from $s^f$ to $d^f$&Continuous - 
			[units]&\eqref{eq3}, \eqref{eq:cost}\\
			\hline
		\end{tabular}
	}
\end{table*}

\begin{review}
	In this section, we formalize the traffic engineering with QoS requirements 
	as a max-min fair allocation problem. Then, we define an objective function 
	to proactively find paths for flows over multiple rounds. Finally, we give 
	an example for the problem before and after a link failure. 
	Table~\ref{tab:tab1} presents the main notation used in the paper.
\end{review}

\subsection{Link Capacity, Flow Conservation, Delay, and 
Cost}\label{sec:QoSRoutingDelay} 
We assume that we have bi-directed graph $G=(\mathcal{S},\mathcal{E})$ where 
$\mathcal{S}$ is a set of P4 switches which are connected to each other (where, 
$|\mathcal{S}|\triangleq S$) and $\mathcal{E}$ is a set of edges where 
$|\mathcal{E}|\triangleq E$. Also, we can transfer a set of flows between two 
pairs of switches.

\noindent{\textbf{Link capacity.}} Equation~\eqref{eq1} ensures the link 
capacity between each pair of P4 switches ($n$ and $m$). Let $f$ be a single 
flow crossing link $n$ and $m$, $\forall f\in \mathcal{F}$.    
\begin{align}\small 
&\sum^{|\mathcal{F}|}_{f=1}{\left(\Phi_{(n,m)}^f \ \cdot \ R^f\right)}\leq \mu 
\cdot  B_{(n,m)},\forall n,m\in \mathcal{S},\label{eq1}
\end{align}
where $R^f$ is the required bandwidth for the $f$-th flow; $\mu$ is a ratio of 
crossing traffic to total bandwidth of each link; $B_{(n,m)}$ is the matrix of 
link bandwidth between $n$ and $m$, and $\Phi_{(n,m)}^f$ is the network 
resource assignment matrix between $n$ and $m$ for the flow $f$.

\noindent{\textbf{Flow conservation.}} Equation~\eqref{eq2} indicates the flow 
conservation and its limitation applied in the presented topology. If a flow 
leaves its source switch $s^f$, then it can not return to the source (no loop-- 
see the first equality). If a flow enters a destination switch $d^f$, it 
remains there (see the second equality). Finally, the total input flows from a 
node should be the same as the total output flows on the same node ($n$) (see 
the third equality).
\begin{align}\small 
&\sum^N_{m=1}{\Phi_{(n,m)}^f}-\sum^N_{m=1}{\Phi^f_{(m,n)}}=
\begin{cases}
1 & \text{if } n=s^f\\
-1 & \text{if } n=d^f\\
0 & \text{Otherwise}\\
\end{cases}.%\nonumber\\
%\forall f\in \mathcal{F},\ \forall n\in\mathcal{S},
\label{eq2}
\end{align}

\noindent{\textbf{Propagation delay.}} 
We can control the propagation delay of each flow within a path $p$ using 
Eq.~\eqref{eq3}. This equation ensures that the total delay of encountered pair 
switches per-flow of a path $p$ should be at most equal to maximum tolerable 
delay for each $p$ or $T^p$. Equation~\eqref{eq3} satisfies loop prevention for 
each flow $f$. This equation ensures that the delay of the path is less than 
the threshold value $T^p$. 
\begin{align}\small 
&\sum^{|p|}_{(n,m)}{\left(\Phi^f_{(n,m)}\ \cdot\ D_{(n,m)}\right)}\leq T^p,\ 
\forall f\in \mathcal{F}, \forall (n,m)\in p,\label{eq3}
\end{align}
where $D_{(n,m)}$ is the propagation delay of link $(n,m)$; $p$ is a path from 
$s^f$ to $d^f$ consisting a set of links between a source and a destination. 
Equation~\eqref{eq4} ensures that each flow crosses each link once.   
\begin{align}\small 
&\sum^N_{m=1}{\Phi_{(n,m)}^f}\leq 1,\ \forall n\in \mathcal{S},\ \forall f\in 
\mathcal{F},\label{eq4} \\
&\Phi_{(n,m)}^f\in \left\{0,\ 1\right\},\ \forall n,m\in \mathcal{S},\ \forall 
f \in \mathcal{F}.\nonumber
\end{align}

\noindent\textbf{Cost.} We model the cost of steering traffic flows from path 
$p$ as follows.
\begin{align}\small 
&C_{(n,m)}=\sum^{|\mathcal{F}|}_{f=1}{C^f_{(n,m)}}, \quad\forall (n,m)\in 
p,\label{eq:cost}
\end{align}
where $C^f_{(n,m)}$ is the cost of flow $f$ while crossing link $(n,m)$.

\subsection{Objective Function}\label{sec:SFC}

Our main objective is 
% to accommodate the maximum number of flows in the network and the minimum 
%associated costs
to allocate the flows according to the available resources using max-min fair 
allocation. Equation~\eqref{eq:objective} calculates this function 
$\phi_{(n,m)}$ which implies the weighted function per link $(n,m)$. 
\begin{align}\small 
&\phi_{(n,m)}=\alpha \cdot \frac{B^u(n,m)}{B(n,m)} +\lambda \cdot 
C_{(n,m)}+\zeta\cdot D_{(n,m)},\ \forall n,m\in \mathcal{S}, 
\label{eq:objective} 
\end{align}
where the fraction $\frac{B^u(n,m)}{B(n,m)}$ is the bandwidth utilization (or 
link utilization) and $C_{(n,m)}$ is the cost of steering traffic from switch 
$n$ to $m$ (it is an input of the problem); $\alpha$, $\lambda$ and $\zeta$ are 
the coefficients and have values between 0 and 1. 
In this way, if the priority is given to the delay, we require to define high 
ratio for $\zeta$. Otherwise, if the link cost is the highest priority we need 
to provide higher ratio for $\lambda$, and finally, if link utilization is the 
importance criterion for the application, we need to have higher value for 
$\alpha$ coefficient in the objective function.
\begin{review2}
	Herein, we state some example use-cases. Many cloud service providers, 
	e.g., Google, Amazon, Microsoft, offer network services with different QoS 
	requirements to the users in different locations~\cite{george-TON14}. For 
	example, network services that offer critical business transactions are 
	sensitive to delay and bandwidth rather than cost. While other services, 
	offered by Content Delivery Network (CDN) providers, that transfer a large 
	volume of data dictate high data transmission 
	cost~\cite{googlecost,micosoftAzure}. Furthermore, in 5G use-cases, we 
	require to provide Ultra-Reliable Low-Latency Communication (URLLC) channel 
	to satisfy mobile users' demands~\cite{5GURLLC}. Therefore, depending on 
	the use-case, the coefficient for the different parameters can be set.   
\end{review2}

Here, we apply a max-min fair allocation~\cite{maxminfair} considering the 
$\phi_{(n,m)}$ for each link. The network administrator can select the related 
coefficient of network latency and bandwidth usage according to the network 
(flow) requirements. Specifically, large flows demand high bandwidth while the 
short flows are sensitive to delays. Therefore, according to the flow sizes, 
desired values for the coefficients of the participant terms 
in~\eqref{eq:objective} are adjusted. The QoS is a priority given to each flow 
according to the application requirements. For example, an application can 
request low transmission delay~\cite{RFC-qos}.

\noindent\textbf{Definition:} Flow $f$ bottlenecks at link $(n,m)$ if the 
following conditions occur: 
\begin{enumerate}
	\item $B^u (n,m)=1$;
	\item Flow $f$ obtains the maximum rate of all flows crossing link $(n,m)$.
\end{enumerate}

\begin{figure}[ht]
	\centering
	\centering
	\includegraphics[width=\columnwidth]{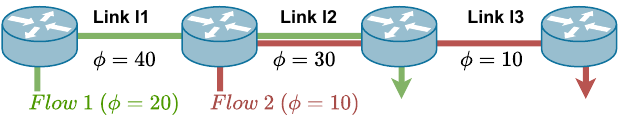}
	\caption{A simple running example network illustrate fairness before 
	failure occurs. Flow 2 (red) is bottelnecked to ${\phi}_{(n,m)=10}$ and 
	Flow 1 (green) is ${\phi}_{(n,m)=20}$.}
	\label{fig:maxminExample}
\end{figure}

\noindent\textbf{Running example:}  Fig.~\ref{fig:maxminExample} presents a 
network without failure with 3 links carrying 2 flows. The max-min fair 
allocation of flow 2 (red) with $\phi=10$, which is bottlenecked at link $l3$, 
and flow 1 (green) with $\phi=20$, which is bottlenecked at link $l2$. Link 
$l2$ has spare $\phi$ since (green) flow 2 is bottlenecked elsewhere. 

% Herein, we first order all the flow requests in ascending order of $\phi$. 
%Then, we start to assign the flow requests from the minimum to maximum request 
%value of $\phi$ in the network.

\begin{figure}[ht]
	\centering
	\centering
	\includegraphics[width=\columnwidth]{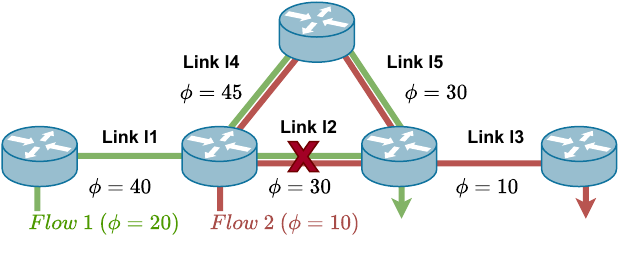}
	\caption{A simple running example network illustrate fairness after failure 
	occurs. Flow 2 (red) is bottelnecked to ${\phi}_{(n,m)=10}$ and Flow 1 
	(green) is ${\phi}_{(n,m)=20}$.}
	\label{fig:maxMinExampleFailure}
\end{figure}

Fig.~\ref{fig:maxMinExampleFailure} illustrates max-min fair allocation in the 
the presence of a link failure ($l2$). The traffic steers from links $l4$ and 
$l5$ after failing to link $l2$ with satisfying the max-min fair allocation.

\section{TEL: The Solution}\label{sec:algo}

In this section, we describe our contribution to failover traffic engineering. After modeling the network requirements, we should gather the network information to find the best paths while preserving the QoS constraints. 
\begin{review}
We first explain \systemnameTEL. Then, we describe \systemnameDC (see Section~\ref{sec:TEL-D}). 
\end{review}
\systemname has three different phases, namely, \textit{network monitoring}, \textit{path selection} and \textit{rule generation}. In the first phase (Section~\ref{sec:monitoring}), the information of the network is collected by the P4 runtime to use it as the input for the path selection phase. We select $K$ unique shortest paths for $K$ network services/applications using the concept of DLA to carry the flows in the path selection phase (see Section~\ref{sec:path-selection}). Finally, \systemname generates a set of proper forwarding rules according to the chosen shortest paths in the previous phase (see Section~\ref{sec:rule-generation}).

\subsection{Network Monitoring}\label{sec:monitoring}
To obtain the network information, we should send a set of probe packets periodically to the network. These packets can be sent either by the controller or end-hosts because there is no packet generation mechanism available in a P4-enabled device.
We use P4 runtime to obtain the information on the network. The controller of TEL uses probe packets to get network information like the propagation delay and bandwidth. Each probe packets contains a set of fields used to collect the link information. The information of links is cloned to the controller to build the network topology. The time to send such probe packets can be tuned depending on the user's needs. However, the path selection phase requires network information before applying the selection procedure. We also assume that the network topology is known in advance to the controller.

\subsection{Path Selection}\label{sec:path-selection}
Selecting an optimal forwarding path with multi-constrained QoS requirements is a well-known NP-hard problem~\cite{qos-routing-jsac96}, and we limit NP-hard discussions here. Instead, we focus on the detail of our approach. We use the concept of DLA, which is a reinforcement learning approach to solve a problem. Each DLA is a network of Learning Automaton (LA). 
\begin{review}
LA is used for learning simple actions using some simple agents. However, its functionality is limited to the purpose of this work while considering several network-related parameters. Interestingly, DLA is composed of a network of LAs and can be applied to complex network problems. Therefore, for the need of this paper, we adopted a DLA solution. Also, the convergence of a solution using DLA was proven by using Martingale Theorem~\cite{meybodi2014extended}. 
\end{review}

In the LA concept, there is a set of actions for each LA to pick at any time, and the LA randomly selects one of them. There is a probability associated with each action. The random environment supplies the reinforcement signal $\beta$ to the chosen action of an LA agent. The LA updates the action probability based on the received signal. This is called the \textit{training phase of LA}, like the other reinforcement learning-based mechanisms. When the learning phase ends, the LA selects the best action among the available actions. To do so, the LA checks the probability of its actions and returns the action's index with the highest probability as the best action.

We create a corresponding DLA graph of the network graph. Each node in the DLA graph has an LA helping it to choose the best action. 
% There is an associated LA with each node in the DLA graph helping the node to choose the best action. 
% This results in the path selection. 
The number of actions for each LA is equal to the number of outgoing edges $\mathcal{O}$ from each node in the network graph. The initial probability of each action is $\frac{1}{\mathcal{O}}$. Selecting an action by the LA of each node corresponds to selecting a neighbor node in the network graph. 
% We use a similar idea of~\cite{rrdla-TIE-19} in the path selection phase of this work as it finds a solution for this problem in linear time. 

\begin{review}
Specifically, in this paper, we utilize variable-structure automata~\cite{Narendra:LA} on a P-model LA ($\beta_i$ is a fixed value: zero or one). We select this automaton because we require to decide if the explored path is better than the previous one (a binary decision).

Let $p(n)$ and $p(n+1)$ be the action probability vectors at the $n^{th}$ and $(n+1)^{th}$ round of learning, respectively. Hence, we define learning algorithm $T$ in Eq.~\eqref{eq-T}:
\begin{equation} \label{eq-T}
p(n+1) = T[p(n),\alpha(n),\beta(n)],
\end{equation}

We define the operation of the LA below. Considering $p(n)$ as an action probability vector, the LA arbitrarily chooses an action $\alpha_i(n)$ and applies it to the environment. When we receive the feedback from the environment, the LA algorithm updates its action probability vector using Eq.~\eqref{eq:la-reward} for $\beta=0$ and \eqref{eq:la-penalty} for $\beta=1$:

\begin{equation} \label{eq:la-reward}
\begin{aligned}
%\begin{split}
& {p_{i} (n+1)=p_{i} (n)+a(1-p_{i} (n)) \quad j=i} \\
& {p_{j} (n+1)=(1-a)p_{j} (n)\quad \forall j, j\ne i}
%\end{split}
\end{aligned}
\end{equation}

\begin{equation}
\label{eq:la-penalty}
\begin{aligned}
%\begin{split}
& {p_{i} (n+1)=(1-b)p_{i} (n) \quad j=i} \\
& {p_{j} (n+1)=\frac{b}{r-1}+(1-b)p_{j} (n)\quad \forall j, j\ne i}
%\end{split}
\end{aligned}
\end{equation}
where $r$ is the number of actions; $p_{i}(n)$ is the probability of action $\alpha_i$; $p_{j}(n)$ is the probability of action $\alpha_j$; $a$ is the reward, and $b$ is the penalty parameters, respectively.
\end{review}

\begingroup
% \removelatexerror% Nullify \@latex@error
\IncMargin{1em}
\begin{algorithm}[!ht]
\SetKwData{Left}{left}\SetKwData{This}{this}\SetKwData{Up}{up}\SetKwFunction{Union}{Union}\SetKwFunction{FindCompress}{FindCompress}\SetKwInOut{Input}{input}\SetKwInOut{Output}{output}
\SetAlgoLined
% \SetKwFunction{FMain}{NegationDetection}
\Input{ \\ -- The set of sources and destinations \\ -- The flow request}
\Output{\\ -- $K$-shortest paths $\{p_1,\ldots,p_k\}\in \mathcal{P}$}
\BlankLine
 read the network graph file $G$\;
create DLA graph from the network graph\;
 equip each node in the DLA graph with LA\;
 $\mathcal{P}_{best}\leftarrow$ The best path in each iteration of DLA\;
 $I_{th} \leftarrow$ The number of iterations for stop condition\;
$\mathcal{P}\leftarrow$ The set of paths\;
 $\mathcal{B}\leftarrow$ The set of backup paths\;
 $L_p\leftarrow$ The list of explored paths for each ($s$, $d$)\;
 $\mathcal{P}_{cur}\leftarrow$ The current best explored path  each ($s$, $d$)\;
 $counter\leftarrow$ 0 \tcc*[l]{A counter for the paths.}
\While{$counter \leq K$}{
 $s,d \leftarrow$ a unique random source and destination\;
 $s \leftarrow$ a random node\;
 $d \leftarrow$ a random node\;
 $val\mathcal{P}_{cur} \leftarrow \emptyset$\;
 $L_p \leftarrow \emptyset$\;
  \Repeat{the stop condition of LA is met ($I_{th}$)}{ \label{alg:start-la}
   \Repeat{$d$ is not visited}{\tcc*[l]{action selection is equal to selecting a neighbor node.}
 $s$ randomly select an action\;
 activate the LA of the corresponding action\;
 $\mathcal{P}_{cur}\leftarrow \mathcal{P} \cup s$\;
 disable the selected action of $s$\ \tcc*[l]{this reduces the search space of problem.}
}   %\Until {$d$ is not visited}
 evaluate the path using objective function in Eq.~\eqref{eq:objective}\label{alg:eval}\;
  \If { $val(\mathcal{P}_{cur}) \le val(\mathcal{P}_{best})$}{
       reward the selected actions in $\mathcal{P}_{cur}$  using~Eq.~\eqref{eq:la-reward}\;
       $\mathcal{P}_{best} \leftarrow \mathcal{P}_{cur} $\label{alg:pbest}\;
       $L_p \leftarrow  \mathcal{P}_{cur} \cup  L_p$\;
      } 
   Enable all the actions\;
   } \label{alg:stop-la}%  \Until{the stop condition of LA is met ($I_{th}$)}
     $ \mathcal{P} \leftarrow \mathcal{P} \cup \mathcal{P}_{best}$\;
    $\mathcal{B} \leftarrow \mathcal{B} \cup L_p[1]$\;
    $counter\leftarrow counter+1$\;
  }
 \caption{The delay-ranked algorithm}
 \label{alg:path_sel2}
\end{algorithm}\DecMargin{1em}

Algorithm~\ref{alg:path_sel2} presents the pseudo code of our path selection algorithm.
This algorithm runs in several iterations (lines~\ref{alg:start-la}-\ref{alg:stop-la}), and the DLA explores a solution among all the candidate solutions in each iteration. Here, the DLA finds a path from a source $s$ to a destination $d$ from the network graph $G$. At the end of each iteration, the chosen path is examined based on Eq.~\eqref{eq:objective} (see line~\ref{alg:eval}). If the result of evaluating the current path's objective function is better than the previous value, the environment generates a reward for the selected path. 
\begin{review}
Then, all the chosen actions by the LA of each node are rewarded based on~Eq.~\eqref{eq:la-reward} that results in placing all the chosen nodes in $\mathcal{P}_{best}$ as the best-selected path until now (see line~\ref{alg:pbest}). 
% In Eq.~\eqref{eq1-la}, $p_i(t)$ is the probability of action $i$ at time $t$ with $a$ and $b$ as the reward and penalty parameters, respectively.
\end{review}
%  \begin{equation} 
% \begin{flalign}\label{eq:LA}\small 
% \begin{split}
% & {p_{i} (t+1)=p_{i} (t)+a\left(1-p_{i} (t)\right)} \\
% & {p_{j} (t+1)=(1-a)p_{j} (t)\quad \forall j,j\ne i}.
% \end{split}
% \end{flalign}
%  \end{equation}
This procedure continues until the stop condition is met. We define a fixed integer value of $I_{th}$ as the stop condition (see line~\ref{alg:stop-la}). The nodes in the $\mathcal{P}_{best}$ will be chosen as the path for the requested traffic flow. At the end of this phase, Algorithm~\ref{alg:path_update} updates the capacity of the links in the network graph $G$ by applying the requested flow $f$ in the chosen path $p$. 
\begin{review}
After finding a path for each flow request, the network graph should be updated with a new objective function. Thus, this parameter varies over time, and DLA learns to choose the best paths for the flows using the updated network graph.
\end{review}

Running the above procedure results in a path selection. In a provider network which has $K$ different service demands, we need to select a path for each one. %However, we require to choose $K$ different paths from the network graph for different service.
To do so, we repeat the same procedure $K$ times to find a path for each requested service.

\begingroup
% \removelatexerror% Nullify \@latex@error
\IncMargin{1em}
\begin{algorithm}[!bpt]
\SetKwData{Left}{left}\SetKwData{This}{this}\SetKwData{Up}{up}\SetKwFunction{Union}{Union}\SetKwFunction{FindCompress}{FindCompress}\SetKwInOut{Input}{input}\SetKwInOut{Output}{output}\SetKwFunction{proc}{UpdateBandwidth}
\SetAlgoLined
% \SetKwFunction{FMain}{NegationDetection}
% \Input{ \\
%     -- network graph $\mathcal{G}$ \\ -- links delay \\ -- number of workers}
% \Output{\\
% -- a set of nodes $\mathcal{N}_{best}$}
% \BlankLine
\SetKwProg{myproc}{Procedure}{}{}
 \myproc{\proc{$p$,$f$}}{
\For {\textbf{each} $(n,m) \in p$}{
     $C_{(n,m)} \leftarrow\;
    C_{(n,m)}-f$\;
  }\label{alg:stop-nodes}
  }
 \caption{Update path weights}
 \label{alg:path_update}
\end{algorithm}\DecMargin{1em}

\noindent\textbf{Pruning rules.}
We use a Boolean value along with each action of LA to determine if the action can be chosen by the DLA. The LA can select an action \textit{if and only if} the corresponding value to that action is $True$. We apply several pruning rules to speed up the running time of the algorithm and its convergence as follows:
\begin{itemize}[leftmargin=*]
    \item The corresponding actions of the nodes that are selected in each iteration are disabled. This helps in reducing the search space in the DLA graph. At the end of each iteration, all the disabled actions are enabled again to contribute to the next round of the path selection process.
      Disabling the corresponding action of the chosen edge avoids any routing instability. It also helps to improve the convergence of the algorithm.
    \item The DLA removes a node from the currently selected path if there is no possible action selected from that node. This rule prevents dead-end path selection in each round.
    \item We disable the corresponding actions of the links in each LA that do not have spare bandwidth to place other flows. This rule prevents capacity oversaturation.
\end{itemize}

Fig.~\ref{fig:dla} presents the interaction of LA and the random environment in the DLA theory. The adjacency list of each node in the network graph forms the action-list of each LA. We also add the relevant parameters of \systemname to show how they interact. We prune each LA's action list to speed up the convergency of LA to the optimal action. In such cases, if an action of LA is disabled during an iteration, then the probability of that action remains unchanged while the probability of other action updates according to Eq.~\eqref{eq:la-reward}.

\begin{figure}[ht]
\centering
    \centering
    \includegraphics[width=\columnwidth]{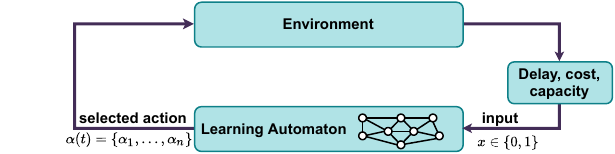}
    \caption{The abstract architecture of learning automaton for \systemnameTEL.}
    \label{fig:dla}
\end{figure}

\noindent\textbf{Time complexity of path selection.} We describe the time complexity of Algorithm~\ref{alg:path_sel2}. Each round of the DLA algorithm requires $O(E)$ to find a path where $E$ is the number of edges in the DLA graph. This procedure repeats $I$ times to find the best path from a source to a destination. Thus, the running time of lines~\ref{alg:start-la} to~\ref{alg:stop-la} is $I \times O(E)$. However, we run this procedure $K$ times to find $K$ different paths. Therefore, the total running time of the path selection phase is $K\times I\times O(E)$.

The above time complexity show that \systemname can solve the problem in a linear time by preserving the QoS constraints for different network applications.

\subsection{Rule Generation}\label{sec:rule-generation}
The rule generation phase creates a proper set of table entries for the 
corresponding switches (i.e., the selected nodes) in each path. These rules 
should be installed on the switches using the P4 agent to steer the traffic. 
However, to differentiate the traffic of different flows that we use an ID for 
each path (see more detail in Section~\ref{sec:p4-implementation}).

To generate the forwarding rules, we keep track of all paths from the sources to the destinations, including the intermediate nodes. There might be multiple paths available between each pair of sources and destinations, but \systemname exploits the ones explored during the path selections phase. Therefore, to generate each forwarding rule for a given P4 switch, we have to check the proper egress port. Suppose that we have a path from $s1\rightarrow s2 \rightarrow s3$. When we generate the forwarding rule for $s1$, we check the network graph for the egress port of $s1$ that is connected to $s2$. Then, the egress port number of $s1$ along with the source and destination IP addresses of this path is inserted as a forwarding rule into $s1$. We follow the same procedure for $s2$ and $s3$ in this path.

\begin{review}
\subsection{\systemnameDC}\label{sec:TEL-D}
\systemname proactively computes paths, i.e., primary and backup using DLR by checking the network. A simplified version of the path selection algorithm of \systemname without the need for the control plane interaction is called \systemnameDCTEL. \systemnameDC is suitable for datacenter topologies where the servers are within a few hops distance from each other and link latencies are short. Thus, we use the basic idea of path selection in the DLA manner in \systemnameDCTEL. 
\begingroup
% \removelatexerror% Nullify \@latex@error
\IncMargin{1em}
\begin{algorithm}[!ht]
\SetKwData{Left}{left}\SetKwData{This}{this}\SetKwData{Up}{up}\SetKwFunction{Union}{Union}\SetKwFunction{FindCompress}{FindCompress}\SetKwInOut{Input}{input}\SetKwInOut{Output}{output}\SetKwFunction{proc}{findPort}
\SetAlgoLined
% \SetKwFunction{FMain}{NegationDetection}
\Input{port\_status}
\Output{a port}
\BlankLine
\SetKwProg{myproc}{Procedure}{}{}
 \myproc{\proc{}}{
\For {\textbf{each} $port \in port\_status$}{
    \tcc*[l]{We check the corresponding bit of each active port using XOR operator.}
    \If { $port \oplus$ \texttt{0x1==0} }{
     $ActivePorts\leftarrow ActivePorts \bigcup port$;
  }
  }
 nextPort=random($ActivePorts$)\;
  return nextPort\;
  }
 \caption{\systemnameDC in dataplane}
 \label{alg:TEL-DC}
\end{algorithm}\DecMargin{1em}
% \color{black}

\systemnameDC firstly checks all the active ports when the failing occurs to find all the available active ports. We use one bit per port in each switch to set the status of them. By performing a XOR $\oplus$ on the value of each port with \texttt{0b1}, we check if the port is active. Following this operation for all ports, we can list all the available active ports. \systemnameDC randomly selects one of its outgoing egress ports and forwards the flows through that port. 
% This is not the optimal solution but it provides similar performance to those of data plane solutions. 
% This kind of mechanism is more feasible in the datacenter network where the nodes are within a few hops from each other.
Algorithm~\ref{alg:TEL-DC} shows the pseudo code of \systemnameDC in detail.

\end{review}

\section{Proof-Of-Concept}\label{sec:proof}
% \color{blue}
We implement path selection reported in Section~\ref{sec:path-selection} and rule generation (see Section~\ref{sec:rule-generation}) parts of \systemname in Python with around 600 lines of code. We now explain the architecture of \systemname with the PISA switch model (see Section~\ref{TEL-Architecture}). Then, we explain the P4 code implementation (see Section~\ref{sec:p4-implementation}). Finally, we briefly explain the \systemnameDC (see Section~\ref{sec:TEL-D}).

\subsection{\systemname Architecture}\label{TEL-Architecture}

Fig.~\ref{fig:architecture} presents an abstraction of P4 PISA pipeline~\cite{bmv2} with \systemnameTEL. The P4 implementation of \systemname is carried out in P4\_16~\cite{p416} using the BMv2~\cite{bmv2} switch. It contains the control and data plane layers. The control plane is in charge of monitoring the network and selecting the paths according to the network requirements (see Section~\ref{sec:algo}). The data plane forwards the packets based on the forwarding rules generated by the control plane.
The presented model in the data plane includes forwarding pipelines, namely, ingress and egress. In this figure, the network operators can configure the parser to match arbitrary packet header fields. Each pipeline includes a sequence of match-action stages. The ingress and egress pipelines can be programmed using P4 runtime as the control plane agent. 
\begin{figure}[ht]
    \centering
    \includegraphics[width=\columnwidth]{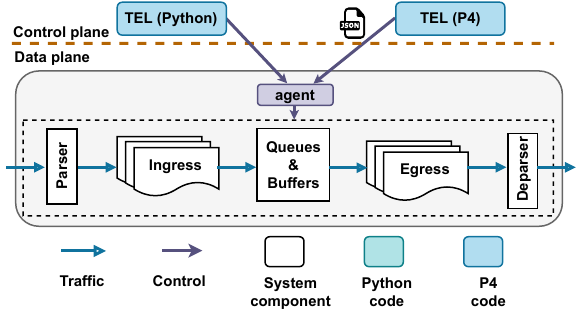}
    \caption{PISA abstraction with \systemnameTEL.}
    \label{fig:architecture}
\end{figure}

We compile the P4 implementation of \systemname and generate the \textit{JSON} representation to load on the switches. In this abstraction model, the forwarding rules for the switches are generated using the Python part of \systemnameTEL.

\begin{figure}[ht]
    \centering
    \includegraphics[width=\columnwidth]{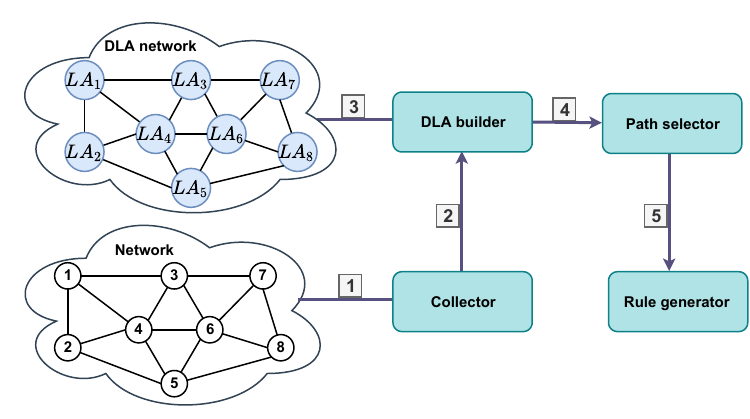}
    \caption{Internal architecture of \systemnameTEL.}
    \label{fig:tel-python}
\end{figure}
Fig.~\ref{fig:tel-python} depicts the internal architecture of \systemname that is implemented in Python. In the beginning, the \texttt{Collector} component monitors the network and obtains the required information (see \fbox{1}). To gather the network information, a set of probe packets are generated. Each switch replies to the probe packets accordingly. Then, the information are fed \fbox{3} into \texttt{DLA builder} component that makes the corresponding DLA graph from the network graph (see \fbox{4}). The \texttt{Path selector} performs path selection using the DLA network and link information. \systemname chooses each path according to the algorithm in Section~\ref{sec:algo}. After computing the paths, we need to generate the forwarding rules through \texttt{Rule generator} component and load them into the switches (see \fbox{5}). We use the P4 local agent for this purpose. The switch is now ready to steer the network traffic.

\subsection{P4 Implementation}
\label{sec:p4-implementation}
To implement our solution in P4, we need to keep the state of all paths in each switch. We use the P4 registers for this purpose by assigning a bit for each path. We call this register \textit{path\_status}, which has 0 value to show a primary path and 1 to indicate the backup one. The overall number of bits is equal to the given number of paths for the topology, i.e., $K$ in Section~\ref{sec:path-selection}. We use $\ceil{\log|\mathcal{P}|}$ bits for this purpose.

We use \textit{two} tables to implement \systemnameTEL. The first table, \textit{table\_1}, matches packets based on the set of source and destination IP addresses that pass a set of links as discussed in Section~\ref{sec:model} and assigns a set of IDs as the \textit{flow\_set} IDs. We use P4 metadata to store the \textit{flow\_set} IDs. The metadata are memory units that can carry packet data within the switch. Each \textit{flow\_set} has $\ceil{\log|\mathcal{P}|}$ bits length. All the incoming flows pass \textit{table\_1} to get the proper ID along with a proper value from the registers for \textit{path\_status}.  
% The \textit{flow\_set} ID generation mechanism of \textit{table\_1} decreases the number of table entries. 

In the second table, \textit{table\_2}, the packets are matched based on the \textit{flow\_set} ID and \textit{path\_status} and forwarded to a proper egress port. We use the basic IPv4 forwarding to forward the traffic of each path. We have \textit{two} set of rules to install on each switch, namely, \textit{primary} and \textit{backup} rules. Both sets are proactively installed. We use the P4 local agent to update the \textit{path\_status} register in the case of any path failure to set the proper values. Fig.~6 depicts the ingress control of \systemname in P4.

% We also use packet recirculation if the port is down, and the \textit{port\_status} register is not updated to avoid the loss of currently processed packets.

\begin{figure}
\begin{tikzpicture}
\node [mybox] (box){%
    \begin{minipage}{0.9\columnwidth}
\begin{lstlisting}
control MyIngress() {
 register<bit<1>>(70) port_reg;
 action host_set(bit<7> new_host) {
        meta.flow_ID=new_host; }
    table table_1 {
        key = {
    			hdr.ipv4.srcAddr: exact;
          hdr.ipv4.dstAddr: exact; }
        actions = {
            host_set; }
    } 
    table table_2 {
        key = {
            meta.flow_ID: exact;
			      meta.port_status: exact; }
        actions = {
            ipv4_forward;  }
    }    
   apply {
      if (hdr.ipv4.isValid()) {
 		   table_1.apply();
       port_reg.read(meta.port_status, (bit<32>)meta.flow_ID);
 		   table_2.apply(); }
     }
}
\end{lstlisting}
    \end{minipage}
};
\node[fancytitle, right=10pt] at (box.north west){};
 \end{tikzpicture}
 \label{fig:tel-p4}
 \caption{Path forwarding of \systemname in P4.}
\end{figure}

\noindent\textbf{Bandwidth monitoring.}\label{sec:Bandwidthmonitoring}
To acquire the link information, we should collect the relevant information and forward them along with probe packets. P4 enables the customized header definition, and we use this feature to monitor the information of links. Therefore, we define the probe packet header, including the number of sent bytes, the last timestamp, and the probe packets' current timestamp. Depending on the number of egress ports in the network, a suitable port for the monitoring information of egress ports can also be defined in the new header fields. The required information of fields for the probe packets is collected by checking the switch's \textit{standard\_metadata} of the switch.

To obtain the available bandwidth information, we need the number of bytes sent since the last probe packet plus the previous and current packets' timestamps. This information is cloned to the controller for the available bandwidth measurements. Afterward, we calculate the link utilization information at the controller. We use P4 registers in each switch to store the number of transmitted bytes and the packet timestamps—the value of these registers updates when a new probe packet enters a switch.

\noindent\textbf{Handling failure. }\label{sec:Handlingfailure}
When a failure occurs on a link, the corresponding bit to that link's status should be set to 0. We use \textit{path\_status} metadata to carry the status of the egress port for the packet. The packets are forwarded according to this metadata's value, either using a primary or backup path.  The packets match with \textit{path\_status} metadata along with the \textit{flow\_set} ID in \textit{table\_2}. Then, the packet is forwarded to the proper egress port accordingly.

\begin{review}
\noindent\textbf{\systemnameDC implementation.}
We implement \systemnameDC in P4 using \textit{two} tables. The first table is used to assign a failure ID (\texttt{FID}) to each failure based on the port status. \systemnameDC uses the second table to forward the packet to an egress port based on \texttt{FID}. For example, if a switch has four ports \texttt{<1,2,3,4>}, then we indicate each port's status with one bit, i.e., 1 to indicate the port is up and 0 to show the port is down. We use \texttt{FID} in the second table to find the \texttt{FWD\_SET} that is a list of candidate active ports to forward the packets. A random number within the list of \texttt{FWD\_SET} is chosen to use as the final egress port to steer the traffic of failed link.

\begin{figure}[!ht]
% \begin{table}
% \renewcommand{\familydefault}{\ttdefault}\normalfont
    \centering
\begin{tabular}{ccc}   % top level tables, with 2 columns
\footnotesize
% caption 1 & caption 2 \\  
% leftmost table of the top level table
\begin{tabular}{|c|c|} 
\hline
\multicolumn{2}{|c|}{\texttt{Table T1}}\\
\hline
{\texttt{port\_status}} & {\texttt{FID}}\\
\hline
\texttt{0111} & \texttt{1}   \\
\texttt{1011} & \texttt{2}  \\
\texttt{1101} & \texttt{3}  \\
\texttt{1110} & \texttt{4}   \\
\hline
\end{tabular} &  % starting rightmost sub table
\small
\begin{tabular}{c} 
\scriptsize
$\rightarrow$
\end{tabular} &
% table 2
\footnotesize
\begin{tabular}{|c|c|c|c|} 
\hline
\multicolumn{3}{|c|}{\texttt{Table T2}}\\
\hline
{\texttt{FID}} & {\texttt{FWD\_set}}& {\texttt{FWD}} \\
\hline
\texttt{1} & \texttt{2,3,4} & \texttt{2}   \\
\texttt{2} & \texttt{1,3,4} & \texttt{1}   \\
\texttt{3} & \texttt{1,2,4} & \texttt{4}   \\
\texttt{4} & \texttt{1,2,3} & \texttt{1}   \\
\hline
\end{tabular} \\
\end{tabular}
    \caption{\text{\systemnameDC FRR encoding.}}
    \label{fig:encodingTable}
% \end{table}
\end{figure}

\end{review}

\begin{review} 
\noindent\textbf{Tofino implementation.} We successfully test \systemname and \systemnameDC on Tofino P4 software switches. Hence, this confirms that the proposed approaches could be applied in real P4 programmable switches. In both approaches, we need to have two tables to install the forwarding rules. Interestingly, we do not need complex match operations in both tables to match the traffic.
\end{review}

\section{Performance Evaluation}\label{sec:results}
\begin{review}
In this section, first, we describe the memory cost of using TEL and testbed. Then, we report the result of comparisons with Yen's K-shortest path algorithm. After that, we compare \systemnameDC against the state-of-the-art FRR mechanism, F10~\cite{F10-NSDI13}, by leveraging circular FRR sequences. Moreover, we compare \systemname with DDC~\cite{DDC-NSDI13} for WAN topologies. Finally, we report the differences between our mechanisms with MPLS and OSPF IP FRRs.
\end{review}

\noindent\textbf{Memory cost.} \systemname uses extra memory to store the backup paths. Considering 25 paths, we use 7 bits for \textit{flow\_set} to encode each path ID and one bit to determine the backup path's usage. We require this encoding to differentiate the traffic of the end-hosts. Otherwise, the traffic could not be forwarded to the right destination. All in all, we need 8 bits in each switch to encode all the paths. We also require the information of the new egress port, i.e., 9 bits, and the MAC address, i.e., 48 bits, for the new path to steer the traffic of the failed path. Therefore, the switches that handle each failure require extra 57 bits for this purpose. 

Each failed path influences the rule update in two switches, and here we explain the reason by providing an example. 
Assume that there are two paths from node 'A' to node 'D', i.e., A$\leftrightarrow$ B$\leftrightarrow$ D and A$\leftrightarrow$ C$\leftrightarrow$ D. If the link (A, B) fails, we need to update the forwarding rules in node 'A' and 'D' to forward the traffic through node 'C'. Thus, nodes 'A' and 'D' require extra 65 bits to handle the failure. 
% \subsection{Discussion} 
% \color{blue}
% \noindent\textbf{Discussion.}
\systemname installs additional forwarding rules on the network devices. To have a resilient and robust system, we should prepare the system for the network changes like a failure. For each link failure, \systemname installs two additional forwarding rules, i.e., one rule in \textit{table\_1} and one rule in \textit{table\_2}.% We claim that this is usual in real-world cloud service applications. If we have an application/user that requires extra services, they need to pay extra costs to have such services.

The memory usage of \systemnameDC depends on the number of ports per switch for \texttt{port\_set} and \texttt{FWD\_SET} fields.

\noindent\textbf{Testbed.}\label{scenarios}
%\subsection{Simulation Scenarios and Setup}\label{scenarios}
% In this part, we separate the scenarios into \textit{two} topologies: \textit{simple} and \textit{complex}. 
We conduct the simulation using Mininet network emulator~\cite{lantz2010network} on an Intel Xeon CPU E5-2667 3.3GH VM with 190 GB RAM and 32 CPU cores running Ubuntu server 18.04. We will make results of \texttt{TEL} fully reproducible in~\cite{tel}.

\subsection{Comparison with K-shortest Path} \label{sec:TELvsKSP}
In this section, we compare the performance of \systemname with Yen's K-shortest path algorithm. The Yen's K-shortest path algorithm finds $K$ shortest paths between a source and a destination. This algorithm's time complexity is $O(KN(M+N\log N))$ where $K$ is the number of calls, $M$ is the number of edges, and $N$ is the number of vertices in the graph, respectively. 

\begin{review2}
\noindent\textbf{K-shortest path algorithm implementation.} To implement Yen's K-Shortest Path (KSP) algorithm, we use Dijkstra's algorithm to find $K$ paths for each source-destination pair. After selecting each path, we update the available bandwidth of the selected links to choose the subsequent paths. Then, we select the best path among the $K$ paths as the primary path. We choose the other paths as a backup when needed. 
\end{review2}

To evaluate the algorithms' performance, we select some pairs of random sources and destinations in Goodnet and AttMpls networks in terms of the average number of hops and running time for various $K$. The value of $K$ varies from 2 and 6 so that in failure scenarios, both algorithms can support up to 6 failed paths. Fig.~\ref{fig:TELvsKSP} shows both algorithms explore paths with a similar number of hops. However, \systemname needs around an order of magnitude less time to find paths when $K=6$. \systemname prunes the search space when looking for a path, which significantly reduces the running time.
\begin{figure}[ht]
	\centering
	\begin{subfigure}{0.98\linewidth}
		\includegraphics[width=\linewidth]{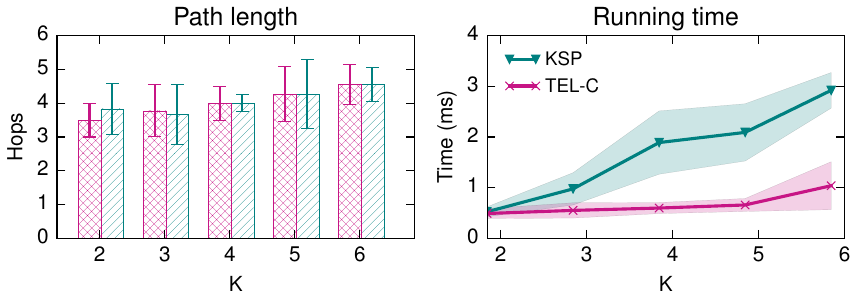}
		\caption{Goodnet}
		\label{fig:KSP-TEL-Goodnet}
	\end{subfigure}
	
	\begin{subfigure}{0.98\linewidth}
		\includegraphics[width=\linewidth]{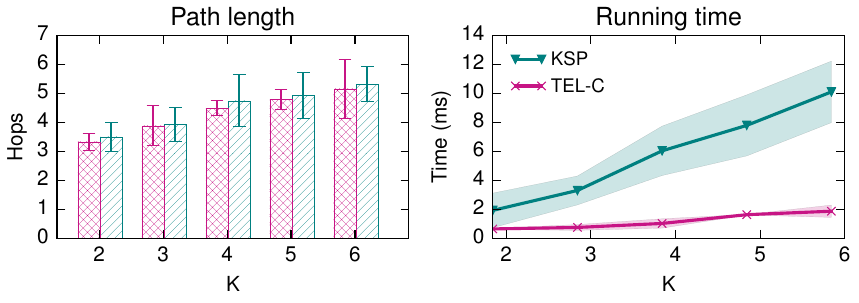}
		\caption{AttMpls}
		\label{fig:KSP-TEL-AttMpls}
	\end{subfigure}
	\caption{The performance of \systemname vs. Yen's K-shortest path algorithm on a) Goodnet and b) AttMpls networks. }
	\label{fig:TELvsKSP}
\end{figure}

\begin{review2}
\noindent\textbf{Quantification of link load:} 
To measure the routing performance of the algorithms, we assess the throughput of each link when every possible link in the network fails. To do so, we take the maximum link utilization over each link failure for all links. We compute the load of individual links $\mathcal{L}_{(n,m)}$ as follows.
\begin{equation}
    \mathcal{L}_{(n,m)} =\max_{\forall (n,m) \in \mathcal{E}}B^u_{(n,m)}.
\end{equation}
We compute this measurement for various failure scenarios per pair of sources and destinations. Furthermore, to satisfy this requirement, the network graphs should be 2-connected. Therefore, we report the results for 2-connected network topologies.

% We compute the relative link
% load $\mathcal{L}_{(n,m)}$ of a specific link $(n,m)\in \mathcal{E}$ by summing up the rates of all traffic aggregates 
% over $(n,m)$. Then we divide the aggregated traffic over the capacity of link $(n,m)$, i.e., $B^u_{(n,m)}$.

% \begin{figure}
%         \centering
%         
%\includegraphics[width=.9\linewidth]{figures/KSP-TEL-trafficLoad-COST239.pdf}
%         \caption{The relative load on each link for failure free case, \systemname, and KSP on COST239 network.}
%         \label{fig:KSP-TEL-trafficLoad}
% \end{figure}
\begin{figure}[ht]
	\centering
	\begin{subfigure}{0.49\linewidth}
		\includegraphics[width=\linewidth]{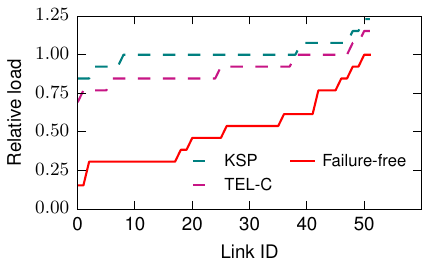}
		\caption{COST239}
		\label{fig:KSP-TEL-trafficLoad-COST239}
	\end{subfigure}
	\begin{subfigure}{0.49\linewidth}
		\includegraphics[width=\linewidth]{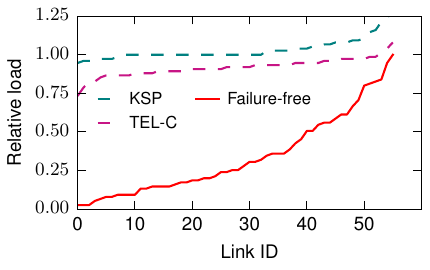}
		\caption{AttMpls}
		\label{fig:KSP-TEL-trafficLoad-AttMpls}
	\end{subfigure}
	\caption{The maximum load on each link for failure free case, \systemname, and KSP where~\ref{fig:KSP-TEL-trafficLoad-COST239}) COST239, and~\ref{fig:KSP-TEL-trafficLoad-AttMpls}) AttMpls networks.}
	\label{fig:KSP-TEL-trafficLoad}
\end{figure}

Fig.~\ref{fig:KSP-TEL-trafficLoad} shows the maximum link load on each link for COST239~\cite{COST239} and AttMpls networks for failure-free, \systemname, and KSP algorithms. We replace the Goodnet network with COST239 since the Goodnet topology is not 2-connected, and we cannot run the link load experiment for our scenarios on this network. We order the link IDs in this figure for presentation purposes. The general trend reports that the maximum load of each link increases when the network faces a link failure. For some scenarios, we see that the maximum link load is over 100\% with \systemname and KSP due to the theoretical analysis, i.e., we consider theoretical link utilizations without packet drops. Furthermore, we observe that \systemname better distributes the load of the network for various failure scenarios.

\end{review2}

\subsection{Performance on Datacenter Topology}
In this part, we compare \systemnameDC with F10~\cite{F10-NSDI13} as the state-of-the-art FRR mechanism. We use leaf-spine topology for comparison with 4 leaf- and spine switches (see Fig.~\ref{fig:DC-topo}). Each link is 100 Mbps. Each leaf switch is connected to 4 servers. %The maximum packet size is 1,300 Bytes. 
Since we are using the Mininet emulator, no additional link delays are applied between switches. We also use the recommendation in~\cite{bmv2-tuning} to achieve high bandwidth throughput of BMv2. 
\begin{figure}[ht]
    \centering
    \includegraphics[width=.95\columnwidth]{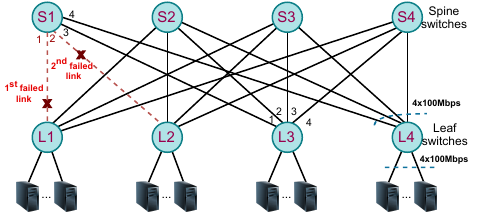}
    \caption{Topology used for emulated evaluation.}
    \label{fig:DC-topo}
\end{figure}

\begin{figure*}[ht]
\centering
 \begin{subfigure}{.24\linewidth}
  \includegraphics[width=\linewidth]{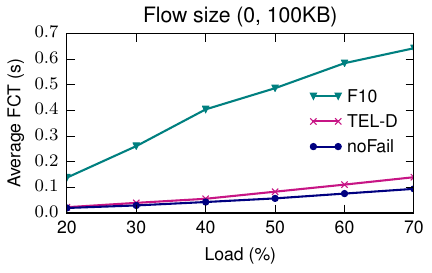}
  \caption{Web-search, 1 link failure.}
  \label{fig:F10-TEL-WebSearch-avgFCT-F1}
 \end{subfigure}
   \begin{subfigure}{.24\linewidth}
  \includegraphics[width=\linewidth]{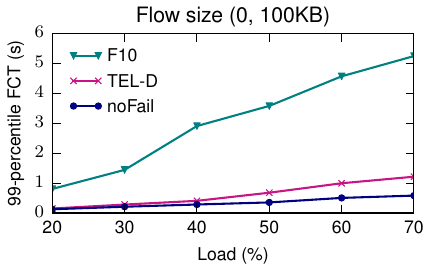}
  \caption{Web-search, 1 link failure.}
  \label{fig:F10-TEL-WebSearch-FCT99-F1}
 \end{subfigure}
 \begin{subfigure}{.24\linewidth}
  \includegraphics[width=\linewidth]{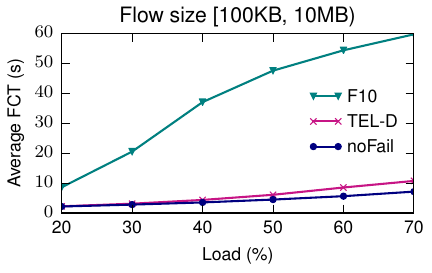}
  \caption{Web-search, 1 link failure.}
  \label{fig:F10-TEL-WebSearch-avgFCTMidSize-F1}
 \end{subfigure}
  \begin{subfigure}{.24\linewidth}
  \includegraphics[width=\linewidth]{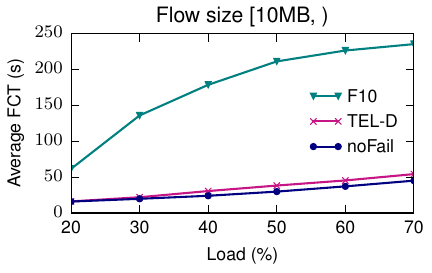}
  \caption{Web-search, 1 link failure.}
  \label{fig:F10-TEL-WebSearch-avgFCTLarge-F1}
 \end{subfigure}
 
 \begin{subfigure}{.24\linewidth}
  \includegraphics[width=\linewidth]{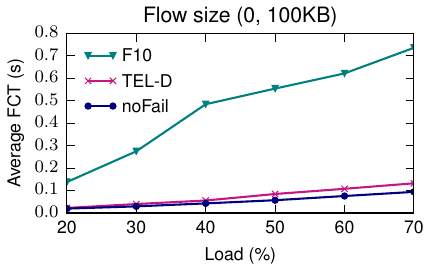}
  \caption{Web-search, 2 link failures.}
  \label{fig:F10-TEL-WebSearch-avgFCT-F2}
 \end{subfigure}
   \begin{subfigure}{.24\linewidth}
  \includegraphics[width=\linewidth]{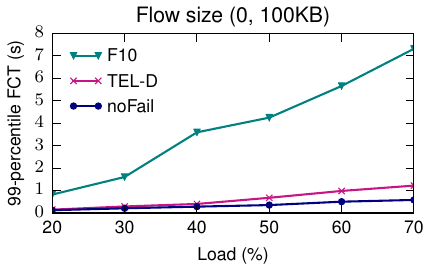}
  \caption{Web-search, 2 link failures.}
  \label{fig:F10-TEL-WebSearch-FCT99-F2}
 \end{subfigure}
 \begin{subfigure}{.24\linewidth}
  \includegraphics[width=\linewidth]{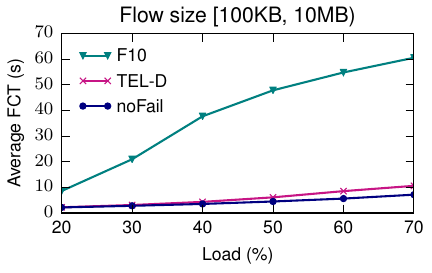}
  \caption{Web-search, 2 link failures.}
  \label{fig:F10-TEL-WebSearch-avgFCTMidSize-F2}
 \end{subfigure}
  \begin{subfigure}{.24\linewidth}
  \includegraphics[width=\linewidth]{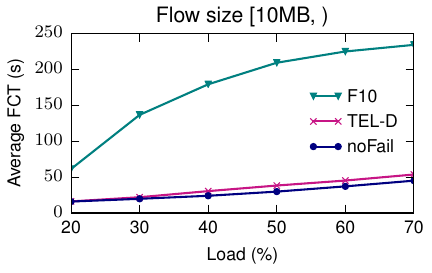}
  \caption{Web-search, 2 link failures.}
  \label{fig:F10-TEL-WebSearch-avgFCTLarge-F2}
 \end{subfigure}
 \caption{Comparison between \systemname and F10 recirculation FRR primitives under 1 and 2 link failures.}
 \label{fig:TEL-F10-WebSearch}
\end{figure*}

\noindent\textbf{F10 implementation in P4.} F10~\cite{F10-NSDI13} is one of the state-of-the-art FRR mechanisms that we implement in our datacenter topology. Considering a datacenter topology with a set of spine and leaf switches with $z$ links between them, F10 can tolerate up to $z$-1 link failures. It also guarantees loop-free packet forwarding. F10 circularly routes the packets. For example, in our topology in Fig.~\ref{fig:DC-topo}, when links (S1, L1) and (S1, L2) fail, F10 forwards packets through port 3 of S1, which is the next available port in the circular sequence. In this example, the packets of failed links are sent to L3, and then we apply FRR on L3, and the packets reach node S4. From there, they will be forwarded to the right destination.

\noindent\textbf{Workloads.} We use the two most popular empirically-derived realistic workloads, i.e., web-search~\cite{alizadeh2010data} and cache-follower~\cite{CacheFollower-sigcomm15}. 
%The reason of using datasets is that since the bandwidth of the links low, we are unable to precisely measure the FCT of large flows ($\geq 10$ MB). 
In either workloads, the distributions of the traffic are heavy-tailed. %Specifically, the data-mining workload is more skewed~\cite{PURR-conext2019} that creates higher imbalances when using ECMP. 
We use the traffic generator in~\cite{bai2016enabling} to generate the desired flows in the network according to a Poisson distribution of each workload and the network load. The load varies in the range 20\% and 70\%. The traffic generator generates different flow sizes as follows. Small flows have the size $<$ 100 KB, while mid-size flows are [100 KB, 10 MB). Large flows have the size $\geq$ 10 MB. We send 500 flows, and the results are averaged over 10 different runs.

\noindent\textbf{Routing and congestion control.} \systemnameDC in a datacenter topology relies on ECMP for load balancing which splits traffic using hash-based mechanism. In this mechanism, the incoming traffic to each leaf switch is randomly forwarded to a spine switch. The spine switch forwards the traffic to the destination leaf switch.

\noindent\textbf{FCT of flows in web-search workload.} Focusing on small flows, \systemnameDC significantly improves the FCT of small flows in the datacenter topology. The results of our experiments for web-search workload were shown in Fig~\ref{fig:TEL-F10-WebSearch}. When one link fails in the topology, \systemnameDC improves the average FCT of small flows between 4.6x to 7.3x compared to F10 in Fig.~\ref{fig:F10-TEL-WebSearch-avgFCT-F1}. \systemnameDC improves 4.3x to 7.1x the 99-percentile FCT of small flows when the topology faces a link failure in Fig.~\ref{fig:F10-TEL-WebSearch-FCT99-F1}. Focusing on mid-size flows, \systemnameDC enhances the average FCT between 3.8x to 7.8x compared to F10 in Fig.~\ref{fig:F10-TEL-WebSearch-avgFCTMidSize-F1}. Focusing on large flows, \systemnameDC improves the average FCT between 3.8x to 6.2x compared to F10 in Fig.~\ref{fig:F10-TEL-WebSearch-avgFCTLarge-F1}. Similar trends exist for 2 link failures. Interestingly, \systemnameDC performs very close to noFail case, especially for the scenarios with load $\leq$ 40\%, because \systemnameDC can select one random egress port among the available active ports to steer the traffic.

\noindent\textbf{FCT of flows in cache-follower workload.} Focusing on small flows, \systemnameDC significantly improves the FCT of small flows in the datacenter topology. The results of our experiments for cache-follower workload were shown in Fig.~\ref{fig:TEL-F10-CacheFollower}. When one link fails in the topology, \systemnameDC improves the average FCT of small flows between 4x to 11.3x compared to F10 in Fig.~\ref{fig:F10-TEL-CacheFollower-avgFCT-F1}. \systemnameDC improves 3.6x to 9.7x the 99-percentile FCT of small flows when the topology faces a link failure in Fig.~\ref{fig:F10-TEL-CacheFollower-FCT99-F1}. Focusing on mid-size flows, \systemnameDC enhances the average FCT between 3.1x to 12x compared to F10 in Fig.~\ref{fig:F10-TEL-CacheFollower-avgFCTMidSize-F1}. Similar to web-search workload results, all the approaches perform similarly in cache-follower workload for 2 link failures. We observe that \systemnameDC performs very close to noFail case in cache-follower workload. Note that since we run the experiments for 500 flows, we do not have large flows results in Fig.~\ref{fig:TEL-F10-CacheFollower}.

\begin{figure*}[ht]
\centering
 \begin{subfigure}{.32\linewidth}
  \includegraphics[width=\linewidth]{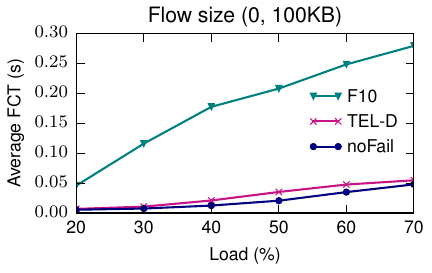}
  \caption{Cache-follower, 1 link failure.}
  \label{fig:F10-TEL-CacheFollower-avgFCT-F1}
 \end{subfigure}
   \begin{subfigure}{.32\linewidth}
  \includegraphics[width=\linewidth]{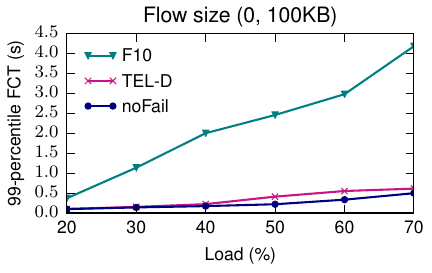}
  \caption{Cache-follower, 1 link failure.}
  \label{fig:F10-TEL-CacheFollower-FCT99-F1}
 \end{subfigure}
 \begin{subfigure}{.32\linewidth}
  \includegraphics[width=\linewidth]{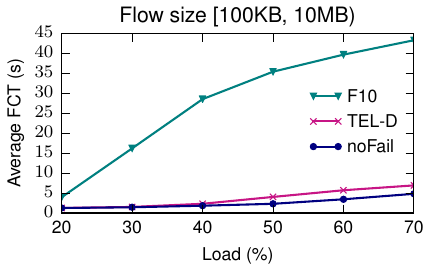}
  \caption{Cache-follower, 1 link failure.}
  \label{fig:F10-TEL-CacheFollower-avgFCTMidSize-F1}
 \end{subfigure}
 
 \begin{subfigure}{.32\linewidth}
  \includegraphics[width=\linewidth]{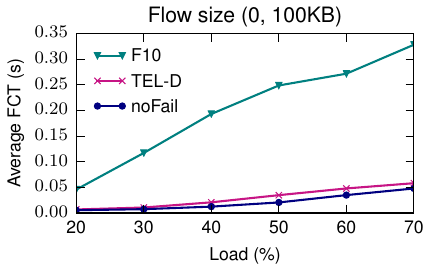}
  \caption{Cache-follower, 2 link failures.}
  \label{fig:F10-TEL-CacheFollower-avgFCT-F2}
 \end{subfigure}
   \begin{subfigure}{.32\linewidth}
  \includegraphics[width=\linewidth]{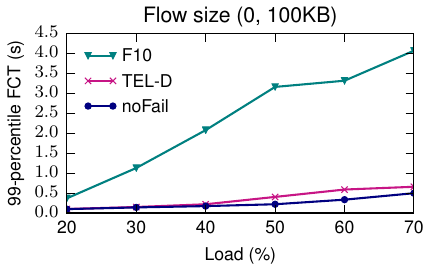}
  \caption{Cache-follower, 2 link failures.}
  \label{fig:F10-TEL-CacheFollower-FCT99-F2}
 \end{subfigure}
 \begin{subfigure}{.32\linewidth}
  \includegraphics[width=\linewidth]{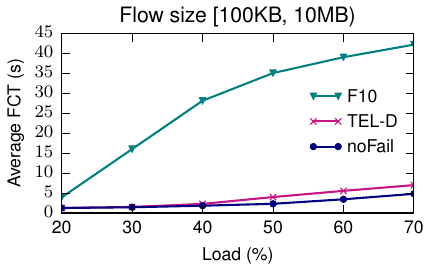}
  \caption{Cache-follower, 2 link failures.}
  \label{fig:F10-TEL-CacheFollower-avgFCTMidSize-F2}
 \end{subfigure}
 \caption{Comparison between \systemname and F10 recirculation FRR primitives under 1 and 2 link failures.}
 \label{fig:TEL-F10-CacheFollower}
\end{figure*}

\subsection{Performance on WAN topology}
In this section, we measure the performance of \systemname and compare it with DDC~\cite{DDC-NSDI13}. 
DCC relies on Gafni-Bertsekas's link reversal algorithms~\cite{GB-TC81} to provide %in-data-plane
connectivity with two versions, namely, partial and full reversal. We implement full reversal due to its simplicity. DDC keeps the nodes' state, such as the direction of each link, and we use P4 registers for this purpose. It also keeps a state in the packet header to compare the sequence number of arrived packets. We define a one-bit header field for the sequence number our implementation in P4.

Each network topology has a different number of nodes and ports. Depending on the number of egress ports of each node, we need to keep a different number of bits in the registers. This is not feasible by having a unique P4 code for all nodes. Therefore, we template the P4 code of DDC using Python Jinja2. To provide full connectivity among the nodes, we need to keep one DAG per node since GB's algorithm is based on DAG. Currently, the P4 language does not support the register of registers and thus, we rely on a single DAG for the evaluation. Note that when running the network topology using Mininet, we load node-specific P4 code on each node.

We use path stretch for comparison, which is defined as the ratio between the length of the path taken by a specific algorithm, i.e., \systemname and DCC, and the shortest path in the current network. Path stretch is affected by the topology, the choice of source and destination, and the number of failed links~\cite{DDC-NSDI13}.

\begin{figure}[ht]
    \centering
    \includegraphics[width=\columnwidth]{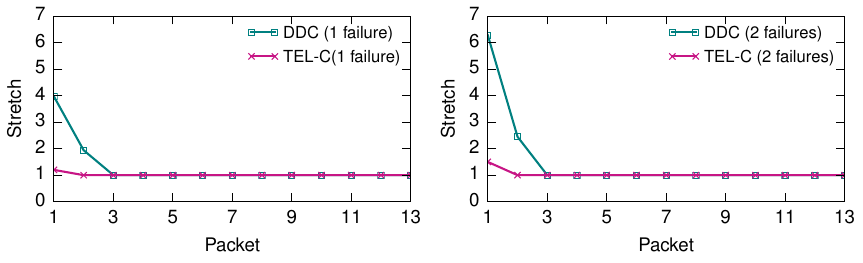}
    \caption{The 99$^{th}$ percentile stretch for AS3967: \systemname vs. DDC for 1 and 2 link failures.}
    \label{fig:DDC-TEL-AS3967-Stretch}
\end{figure}
We compare \systemname and DDC on different topologies such as those of RocketFuel~\cite{Rocketfuel-sigcomm02} similar to DCC paper, and report some representative results. Fig.~\ref{fig:DDC-TEL-AS3967-Stretch} shows the 99$^{th}$ percentile stretch of \systemname and DDC on AS3967 for first 13 packets after link failures. Both algorithms find routes around the failed links after some packets. Generally, by increasing the packets, the stretch decreases to some points, and more link failures result in a higher stretch. \systemname finds routes for failed links faster than DDC. We also checked the median value of stretch for the 13 packets. Both \systemname and DDC have a median value of 1 in Fig.~\ref{fig:DDC-TEL-AS3967-Stretch}.

\begin{figure}[ht]
    \centering
    \includegraphics[width=\columnwidth]{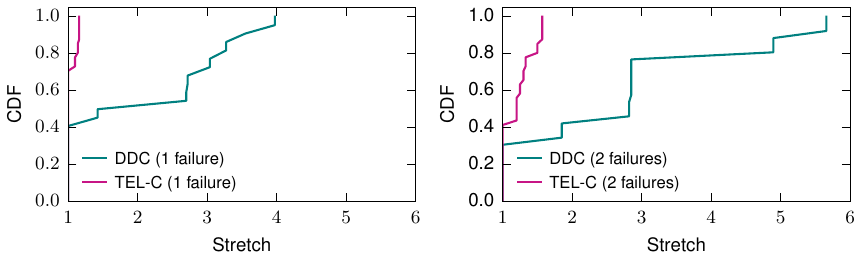}
    \caption{The 99$^{th}$ percentile CDF of steady state stretch for \systemname vs. DDC for 1 and 2 link failures.}
    \label{fig:DDC-TEL-AS3967-CDF}
\end{figure}
We now check the steady-state stretch of the chosen path. Fig.~\ref{fig:DDC-TEL-AS3967-CDF} report the 99$^{th}$ percentile cumulative distribution function (CDF) of stretch for AS3967 with 1 and 2 link failures. Generally, more link failures result in a higher stretch for both algorithms. Since \systemname leverages the idea of backup paths, the stretch is smaller than those of DDC in both 1 and 2 link failures.

\begin{review}
\subsection{Comparison with MPLS FRR}
Multiprotocol Label Switching (MPLS) is widely used to provide FRR in the order of a few 10s of milliseconds delay to repair label-switched path (LSP) tunnels~\cite{mpls-FRR2005}. Similar to \systemname, the backup LSPs are proactively generated to the desired failures. Therefore, the router detours the traffic from the failed port to an alternative by swapping the packet's label on the MPLS stack. The packets follow the new paths until reaching the merge point. At this point, the label of the packets swaps from the backup to primary LSP. However, Resource Reservation Protocol-Traffic Engineering (RSVP-TE) signals establishing backup LSPs among routers.

Unlike MPLS FRR, \systemname does not require a complex signaling mechanism to swap to backup paths. Instead, the controller in the control plane has an overview of the network and provides the required forwarding rules for the backup paths. There is also no need to carry extra labels in the network for the traffic of failed paths.

\subsection{Comparison with OSFP IP FRR}
OSFP IP FRR reduces the reaction time to failure into a few 10s of milliseconds by proactively computing alternative paths. If the primary path fails, then it can rapidly switch to the backup path when the failure is detected. Moreover, OSFP IP FRR requires signal events to its neighbor by using Interior Gateway Protocol (IGP) to recompute the paths for all affected prefixes. Additionally, OSPF Loop-Free Alternate (LFA) FRR can provide a different level of failure protection depending on the network topology.

Unlike OSFP IP FRR, \systemname does not require a signaling mechanism to swap to backup paths. Instead, the controller in the control plane has an overview of the network and updates the required forwarding rules for the backup paths. Interestingly, \systemname allows the selection of more factors in path selection.

\end{review}

% \vspace{-10px}
\section{Applying on Practical Applications}\label{sec:future}

Modern networking applications demand ultra-low-latency delay, and link failure can cause many issues. There have been several network issues during recent years leading to wide Internet outages in different continents such as Asia~\cite{googlefriday}. These kinds of outages result in losing hundreds of thousands of dollars for Google~\cite{googlecost}, affecting thousands of British Airways airline passengers~\cite{BritishAirways}, or disrupting the emergency network~\cite{googleATT}.

Each small delay in many networking applications can lead to a significant drop in business. For example, Akamai in 2017 reported that every 100 milliseconds of delay have a determinant impact in dropping the customers of online businesses~\cite{Akami}. Other networking applications like voice have around 150 milliseconds of tolerable delay, while for gaming applications, this is about 80 milliseconds~\cite{IETF}.

We now explain another practical scenario for big data applications. Distributed stream processing systems like Apache Flink~\cite{Flink} receive data from many resources such as the Internet of Things (IoT) devices, user clicks, and financial transactions. The intermediate results of running a query in such systems should be transferred to the central locations for decision making. The underlying network may fail due to link failure and the highly time-sensitive data require to be rerouted. In such applications, each millisecond of delay is essential for decision making. 

In all the above application scenarios, the failure in delivering traffic can lead to the loss of massive revenues, and TEL can be used in any application scenarios that demand low-latency traffic engineering.

\section{Related Work}\label{sec:relatedwork}
 \begin{table*}[!tpb]
\centering
\caption{Comparison of related works.}\label{tab:relatedcomp}
\resizebox{\textwidth}{!}{
%\begin{adjustbox}{max width=\textwidth}
\begin{tabular}{ccccccc}
\hline
% \coldscr&\coldscr\textbf{Link}&\coldscr\textbf{Flow}&\coldscr\textbf{Propagation}&\coldscr\textbf{Link}&\coldscr\textbf{Operation} &\multirow{1}{*}{\coldscr\textbf{Tools}}\\
% \multirow{1}{*}{\coldscr\textbf{Reference}}&\coldscr\textbf{capacity}&\coldscr\textbf{conservation}&\coldscr\textbf{delay}&\coldscr\textbf{cost}& \coldscr\textbf{mode}&\coldscr\\
\multicolumn{1}{|c|}{\coldscr\textbf{Reference}} & \multicolumn{1}{|c|}{\coldscr\textbf{Link capacity}}& \multicolumn{1}{|c|}{\coldscr\textbf{Flow conservation}}& \multicolumn{1}{|c|}{\coldscr\textbf{Propagation delay}}& \multicolumn{1}{|c|}{\coldscr\textbf{Link cost}}& \multicolumn{1}{|c|}{\coldscr\textbf{Operation mode}}& \multicolumn{1}{|c|}{\coldscr\textbf{Tools}}\\
\hline
\textbf{Data Plane}&&&&&&\\
\cite{PURR-conext2019}&$\checkmark$&$\checkmark$&
$\checkmark$&$\times$&Decentralized&NS3/Mininet\\
\cite{qu2019sqr}&$\checkmark$&$\checkmark$&
$\checkmark$&$\checkmark$&Centralized&Mininet\\

\cite{G975}&$\checkmark$&$\times$&
$\times$&$\times$&Decentralized&$\times$\\

\cite{cholda2009recovery}&$\checkmark$&$\times$&$\checkmark$&$\checkmark$&Centralized&$\times$\\

\cite{zhuang2005failure}&$\checkmark$&$\times$&$\checkmark$&$\checkmark$&Centralized&$\times$\\

\cite{Avallone-TNSM20}&$\checkmark$&$\checkmark$&$\checkmark$&$\checkmark$&Decenterilized&NS3/Mininet\\

\cite{katz2010bidirectional}&$\checkmark$&$\checkmark$&$\checkmark$&$\checkmark$&Distributed&Mininet\\

\cite{steinert2010towards}&$\checkmark$&$\times$&$\times$&$\checkmark$&Distributed&Mininet\\

\cite{cascone2017fast}&$\checkmark$&$\times$&$\times$&$\times$&Centralized&Mininet/BMv2\\
\hline
\textbf{TEL}&$\checkmark$&$\checkmark$&$\checkmark$&$\checkmark$&\textbf{Decentralized}&\textbf{Mininet/BMv2}\\
\hline
\textbf{Control Plane}&&&&&&\\

\cite{Std8021D-1998}&$\checkmark$&$\times$&$\times$&$\times$&Centralized&$\times$\\

\cite{elmeleegy2006count}&$\checkmark$&$\times$&$\times$&$\times$&Centralized&$\times$\\
\cite{rosen2001multiprotocol}&$\checkmark$&$\times$&$\times$&$\times$&Centralized&$\times$\\
\cite{LDP}&$\checkmark$&$\times$&$\times$&$\times$&Distributed&$\times$\\
\cite{qiu2019efficient}&$\checkmark$&$\checkmark$&$\checkmark$&$\checkmark$&Centralized&Mininet/OVS\\
\hline
\end{tabular}}
%\end{adjustbox}\vspace{-10px}
\end{table*}
In this section, we give a summary of different types of failures that have been proposed on the data plane (see Section~\ref{sec:data-plane}) and the control plane (see Section~\ref{control-plane}). The failure on the L2 switch can be detected in legacy networks that require at least 20 milliseconds~\cite{RFC7419}. Considering even 20 milliseconds of delay in detecting failure results in losing a considerable amount of traffic while having Tbps of traffic~\cite{RFC7419,RFC8562}. 

\subsection{Data Plane failure algorithms}\label{sec:data-plane}
In the data plane, we can detect the failures by analyzing the control verification flags of TCP/IP protocol of the metadata of each packet. A summary of FRR solutions in the data plane is reported in~\cite{chiesa2020fast}. For example, the work in~\cite{G975} detects the failure by continuously checking the TCP/IP checksum verification and monitors the increment of bit error ratio while decreasing the data rate quality. The work in~\cite{cholda2009recovery} identifies the failure by validating the throughput plunging and increasing data transmission delay. According to~\cite{zhuang2005failure}, finding a failure on the IP and overlay network is categorized as active and passive solutions. In the former solutions, as reported in~\cite{katz2010bidirectional}, they propose a fast failure detection method called \textit{BFD} that achieves based on the live communication between the neighboring nodes. 
\begin{review}
In the latter solutions, such as~\cite{steinert2010towards}, 
\end{review}
the failure state can be detected based on data packet delivery that is given to other nodes. In this case, the neighbors' nodes can check the packet structure and confirm the links and required operations. However, this type of failure detection requires receiving data flow regularly from the neighbor nodes. Also, the authors in~\cite{Avallone-TNSM20} design a DAG-based algorithm to minimize the number of entries required on the SDN switches. Besides, it decreases the local restoration latency for a failed node/link such that the SDN controller will not be affected. This solution performs only based on the standard features of OpenFlow and avoids inconsistent forwarding tables during updates. The authors in~\cite{cascone2017fast} design SPIDER, a new failure recovery approach that provides a fully programmable abstraction and re-routing policies in SDN. SPIDER aims to minimize the recovery delay and guarantees the failover even when the controller is not reachable. Besides, the work~\cite{borokhovich2014provable} implements a fast failover algorithm in OpenFlow to re-route traffics based on the gathered information from packet headers. This method monitors the packet movement on various routes. It analyzes them based on multiple traversal network graph mechanisms, such as depth-first search and breath-first search. The routing is carried out using failure-carrying packets~\cite{Lakshminarayanan-aigcomm08} algorithm. %Unlike~\cite{cascone2017fast,borokhovich2014provable}, 
\systemnameDC significantly improves the FCT of flows compared to the state-of-the-art FRR mechanisms.
%not only provides a fair allocation and minimizes the delay and capacity but also preserves a failover mechanism on complex network. 

\subsection{Control Plane failure algorithms}\label{control-plane}
Failure faces several routing and data steering issues in SDN, minimizing packet losses and increasing transmission delay. Applying a failure detection mechanism in the control plane leads to having resilient routing in an Ethernet network. In~\cite{Std8021D-1998}, the authors designed a tool based on Spanning Tree Protocol (STP) on the IEEE 802.1D to avoid forwarding loops while providing necessary restoration capabilities. STP also guarantees to establish a unique path between any two nodes. However, it is not equipped to cover failure recovery, and its convergence speed is prolonged up to 50 seconds~\cite{elmeleegy2006count}, which is not an efficient method for real-time applications in large networks.

Some failure recovery solutions are based on MPLS, which can be managed through a data plane. For example, the paper~\cite{rosen2001multiprotocol} utilizes label switching routers to handle the steering packets along with switches by labeling the packet header. They design a label distribution protocol to manage the labeled packets and understand the failure that may happen in the network. Also, the solution's extension is tested and validated on the label distribution protocol reported in~\cite{LDP}. Recently, in~\cite{qiu2019efficient}, the authors design two FRR algorithms managed through a control plane on MPLS. These algorithms can rapidly index the shortest recovery paths and the shortest guaranteed-cost path method to decrease the recovery path's delay cost in an SDN. Unlike the above techniques, \systemname can satisfy the max-min fair allocation.
Table~\ref{tab:relatedcomp} presents a comparison of approaches. The goal of the first category is to present the features of solutions applying in the data plane while the second category does the same for the control plane. Also, the symbol "$\checkmark$" indicates that the approach supports the property; Otherwise, we used "$\times$". Besides, we classify the operational mode into centralized, decentralized, and distributed. 

\section{Conclusion}\label{sec:conclusion}
\begin{review}
This paper presents \textit{two} FRR mechanisms for programmable data plane to steer the traffic with a low failover latency in the failure scenarios. We propose one \textit{proactive} and one \textit{reactive} FRR mechanisms. The first one, \systemnameTEL, calculates the primary and backup paths in the control plane satisfying max-min fair allocation and insert the forwarding rules into the network devices. When failure occurs, the network device can reroute the traffic according to the backup paths. The second one, \systemnameDCTEL, reroutes the traffic in the data plane, making it suitable for self-driving programmable networks. In the future, we plan to extend the \systemname by considering sophisticated traffic policies like priority-based traffic engineering. Also, we plan to extend the \systemnameDC for load balancing scenarios. Furthermore, we plan to extend our learning algorithm with many robust ones such as the max-logit and b-logit~\cite{yaro-2020}.
\end{review}

\section*{Acknowledgment}

This work was funded by the German Ministry for Education and Research as 
BIFOLD - Berlin Institute for the Foundations of Learning and Data (ref. 
01IS18025A and 01IS18037A). Mohammad Shojafar is supported by Marie Curie 
Global Fellowship funded by European Commission with grant agreement 
MSCA-IF-GF-839255.

\bibliographystyle{IEEEtran}
\bibliography{paper.bib}

\begin{IEEEbiography}[{\includegraphics[width=1in,height=1.25in,clip,keepaspectratio]{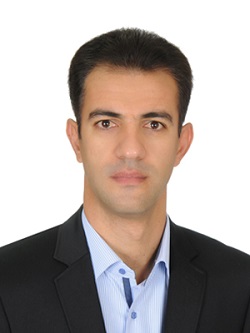}}]{Habib
 Mostafaei} is currently a postdoctoral researcher at the Internet Network 
Architectures (INET) of Technische Universität Berlin. He received a Ph.D. in 
Computer Science and Engineering from Roma Tre University in 2019. He 
currently 
works as a senior researcher in the BIFOLD-BBDC project on networking problems 
for big data analytics. In 2018, he spent eight months as a visiting 
researcher 
at the University of Tuebingen. Before the Ph.D. education, he worked as a 
full‐time faculty member at the Computer Engineering Department of Azad 
University (2009‐2015).  Currently, his main research fields include networked 
systems, network management, and distributed systems. For additional 
information: \url{https://inet.tu-berlin.de/~habib} 
\end{IEEEbiography}
 %\vspace{-35px}

\begin{IEEEbiography}[{\includegraphics[width=1in,height=1.25in,clip,keepaspectratio]{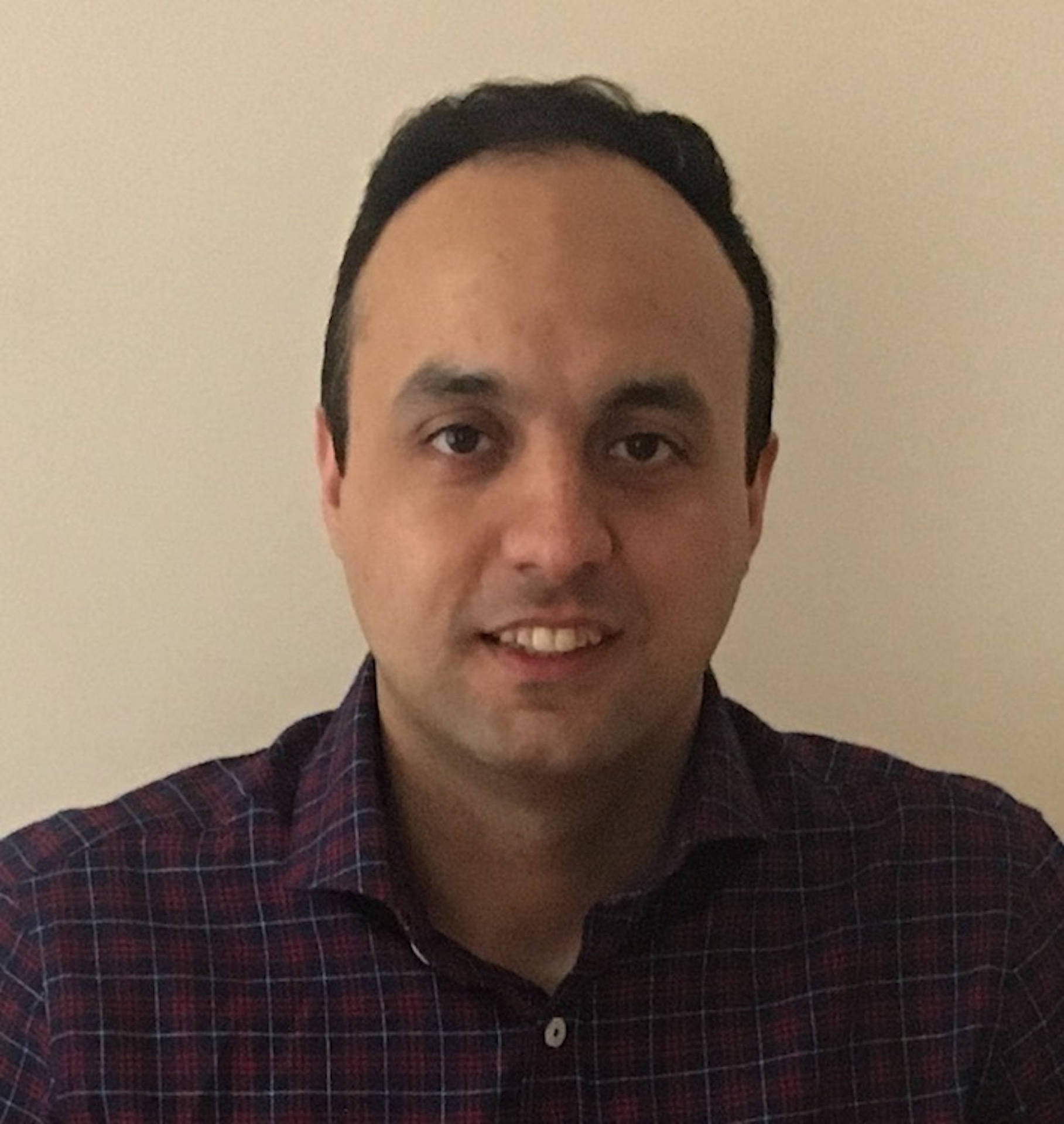}}]{Mohammad
 Shojafar} \textbf{(M'17-SM'19)} is a Senior Lecturer (Associate Professor) in 
the network security and an Intel Innovator, and a Marie Curie Alumni, working 
in the 5G Innovation Centre (5GIC) at the University of Surrey, UK. Before 
joining 5GIC, he was a senior researcher and a Marie Curie Fellow in the 
SPRITZ 
Security and Privacy Research group at the University of Padua, Italy. Also, 
he 
was a CNIT senior researcher at the University of Rome Tor Vergata contributed 
to the 5G PPP European H2020 ``SUPERFLUIDITY'' project. Dr. Mohammad was a PI 
of the PRISENODE project, a 275k euro Horizon 2020 Marie Curie global 
fellowship project in the areas of Fog/Cloud security collaborating at the 
University of Padua. He also was a PI on an Italian SDN security and privacy 
project (60k euro) supported by the University of Padua in 2018 and a Co-PI on 
an Ecuadorian-British project on IoT and Industry 4.0 resource allocation (20k 
dollars) in 2020. He was contributed to some Italian projects in 
telecommunications like GAUChO, SAMMClouds, and SC2. He received his Ph.D. 
degree from Sapienza University of Rome, Rome, Italy, in 2016 with an 
``Excellent'' degree. He is an Associate Editor in IEEE Transactions on 
Consumer Electronics, IEEE Systems Journal and IET Communications. For 
additional information: 
\url{https://www.surrey.ac.uk/people/mohammad-shojafar} 
\end{IEEEbiography}
  %\vspace{-35px}

\begin{IEEEbiography}[{\includegraphics[width=1in,height=1.25in,clip,keepaspectratio]{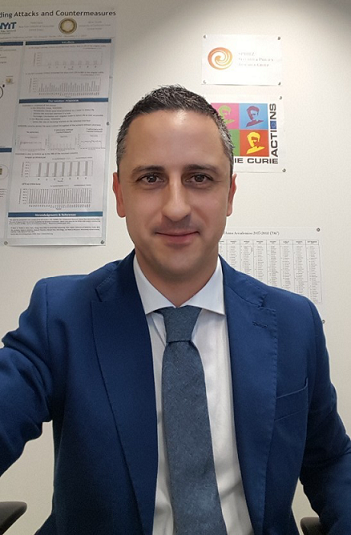}}]
  {Mauro Conti} is Full Professor at the University of Padua, Italy. He is 
also 
 affiliated with TU Delft and University of Washington, Seattle. He obtained 
 his Ph.D. from Sapienza University of Rome, Italy, in 2009. After his Ph.D., 
 he was a Post-Doc Researcher at Vrije Universiteit Amsterdam, The 
Netherlands. 
 In 2011 he joined as Assistant Professor the University of Padua, where he 
 became Associate Professor in 2015, and Full Professor in 2018. He has been 
 Visiting Researcher at GMU, UCLA, UCI, TU Darmstadt, UF, and FIU. He has been 
 awarded with a Marie Curie Fellowship (2012) by the European Commission, and 
 with a Fellowship by the German DAAD (2013). His research is also funded by 
 companies, including Cisco, Intel, and Huawei. His main research interest is 
 in the area of Security and Privacy. In this area, he published more than 350 
 papers in topmost international peer-reviewed journals and conferences. He is 
 Area Editor-in-Chief for IEEE Communications Surveys \& Tutorials, and 
 Associate Editor for several journals, including IEEE Communications Surveys 
 \& Tutorials, IEEE Transactions on Dependable and Secure Computing, IEEE 
 Transactions on Information Forensics and Security, and IEEE Transactions on 
 Network and Service Management. He was Program Chair for TRUST 2015, ICISS 
 2016, WiSec 2017, ACNS 2020, and General Chair for SecureComm 2012, SACMAT 
 2013, CANS 2021, and ACNS 2022. He is Senior Member of the IEEE and ACM. He 
is 
 member of the Blockchain Expert Panel of the Italian Government. He is Fellow 
 of the Young Academy of Europe. For additional information: 
 \url{http://www.math.unipd.it/~conti/}.
  \end{IEEEbiography}
 % \vspace{-35px}.
 \balance

\end{document}